\documentclass[12pt, draftclsnofoot, onecolumn]{IEEEtran}
\usepackage{graphicx}  
\usepackage{epstopdf}
\usepackage{epsfig}
\usepackage{subfigure}
\usepackage[cmex10]{amsmath}
\usepackage{amssymb}   
\usepackage{amsmath}
\usepackage{amsfonts}
\usepackage{cite}
\usepackage{color}
\usepackage{url}
\DeclareMathOperator*{\argmax}{argmax}
\newtheorem{myTheo}{Theorem}
\newtheorem{myLem}{Lemma}
\newtheorem{myRem}{Remark}
\newtheorem{myCas}{Case}

\begin{document}
\title{Hybrid Precoding Architecture for Massive Multiuser MIMO with Dissipation: Sub-Connected or Fully-Connected Structures?}
\author{Jingbo~Du,~\emph{Student Member,~IEEE,}
        Wei~Xu,~\emph{Senior Member,~IEEE,}
        Hong~Shen,~\emph{Member,~IEEE,}
        Xiaodai~Dong,~\emph{Senior~Member,~IEEE,}
        and~Chunming~Zhao,~\emph{Member,~IEEE}
	 \thanks{J. Du is with the National Mobile Communications Research Laboratory, Southeast University, Nanjing 210096, China, and also with the State Key Lab of Integrated Services Networks, Xidian University, Xi'an 710126, China (email: dujingbo@seu.edu.cn). H. Shen and C. Zhao are with the National Mobile Communications Research Laboratory, Southeast University, Nanjing 210096, China (email: \{shhseu and cmzhao\}@seu.edu.cn). W. Xu is with the National Mobile Communications Research Laboratory, Southeast University, Nanjing 210096, China, and also with the Department of Electrical and Computer Engineering, University of Victoria, Victoria, BC V8W 3P6, Canada (email: wxu@seu.edu.cn). X. Dong is with the Department of Electrical and Computer Engineering, University of Victoria, Victoria, BC V8W 3P6, Canada (email: xdong@ece.uvic.ca).}
       \thanks{Part of this material was presented at the IEEE Globecom, Dec. 2017. \emph{(Corresponding author: Wei Xu.)}}

        }

\markboth{}%
{}

\maketitle

\begin{abstract}
In this paper, we study the hybrid precoding structures over limited feedback channels for massive multiuser multiple-input multiple-output (MIMO) systems. We focus on the system performance of hybrid precoding under a more realistic hardware network model, particularly, with inevitable dissipation. The effect of quantized analog and digital precoding is characterized. We investigate the spectral efficiencies of two typical hybrid precoding structures, i.e., the sub-connected structure and the fully-connected structure. It is revealed that increasing signal power can compensate the performance loss incurred by quantized analog precoding. In addition, by capturing the nature of the effective channels for hybrid processing, we employ a channel correlation-based codebook and demonstrate that the codebook shows a great advantage over the conventional random vector quantization (RVQ) codebook. It is also discovered that, if the channel correlation-based codebook is utilized, the sub-connected structure always outperforms the fully-connected structure in either massive MIMO or low signal-to-noise ratio (SNR) scenarios; otherwise, the fully-connected structrue achieves better performance. Simulation results under both Rayleigh fading channels and millimeter wave (mmWave) channels verify the conclusions above.

\end{abstract}
\IEEEpeerreviewmaketitle

\section{Introduction}

\IEEEPARstart{M}{assive} multiple-input multiple-output (MIMO) that employs hundreds or even thousands of antennas is deemed as a reliable technique and has attracted wide attentions due to its numerous potential benefits \cite{5595728, 6736761, 7306538, 7558167}. Several advantages can be obtained in massive MIMO with simple linear precoding, such as cancelling out noise, inter-user interference and fast fading \cite{5595728} \cite{7417293}\cite{7018998}, abilities in secure communication \cite{7328729}. However, conventional digital precoding becomes unsuitable for the massive MIMO systems because it requires that each antenna is connected with a dedicated radio frequency (RF) chain including up-converters, digital-to-analog converters, mixers, power amplifiers, etc. The increasing number of RF chains is a catastrophe in practical circuit design because the RF chains are responsible for a large part of power consumption, hardware cost and implementation complexity.

To address the issue, multiple antennas are attached to the same RF chain which leads to the introduction of hybrid analog and digital processing \cite{6847111, 7400949, 7387790}. Among various setups, the tradeoff between complexity and performance was discussed for the fully-connected array \cite{6717211, 6955826, 7564939, 7947159} and the sub-connected array \cite{7010533, 7445130, Lin2015In}. Contrary to high hardware complexity in the fully-connected structure, complexity reduction at the cost of somewhat reduced performance can be achieved in the sub-connected structure \cite{7248552} \cite{7397861}. \cite{7880698} proposed an improved sub-connected structure to acquire similar performance without increased complexity. 
A switch network was proposed in \cite{7370753} to obtain lower complexity and power consumption. Some ideas of modified zero-forcing (ZF) precoding were also introduced to reduce the system complexity \cite{7160780} \cite{6928432}.

Performance deterioration caused by imperfect hardware is an important effect on hybrid processing. Examples include that a large amount of power dividers and combiners, which do not exist in conventional systems, are applied in the circuit design of a hybrid system. The dissipation caused by these components has a severe impact on the transmit power and it should not be neglected. \cite{7539303} discussed hybrid processing systems under a realistic RF model for the fully-connected structure. \cite{7436794} developed nonlinear power consumption models of hybrid precoding systems and compared different structures.

In addition, considering the effect of quantized precoding and channel information feedback makes the analysis more practical as most current circuit elements are digitally-controlled. For single user MIMO, the optimal transmission strategy over frequency selective channels with limited feedback was developed in \cite{7448873}. \cite{zhu2017two} investigated the pilot design and feedback strategy for hybrid precoding architecture. In \cite{7510972} \cite{8254815}, the quantized analog precoding was analyzed in hybrid processing for multiuser massive MIMO systems. For massive MIMO, \cite{7829396} analyzed the performance of a limited feedback massive MIMO system. The quantized hybrid precoding over limited feedback channels was investigated in \cite{7160780} using a random vector quantization (RVQ) codebook. However, the RVQ codebook was originally designed for conventional isotropically distributed small-scale MIMO channels \cite{1715541} \cite{4100151}. In hybrid analog and digital processing systems, the effective channels, however, are always correlated through the analog precoding. Therefore, the RVQ codebook becomes less suitable for the hybrid system as we will show later in this study. Note that this paper is an extension of \cite{8254815} which only considers the quantized analog precoding when deriving the spectral efficiency. In this paper, we extend \cite{8254815} by additionally considering quantized digital precoding. Moreover, we also provide more insights and simulation results both under Rayleigh and millimeter wave (mmWave) channels.

In this paper, we aim to characterize the impact of realistic analog processing network on the performance of different hybrid precoding structures and find effective codebooks to improve the spectral efficiency. The main contributions of this paper are summarized as follows:

1) We introduce two typical circuit configurations of hybrid precoding systems, namely the sub-connected and fully-connected structures. The spectral efficiencies of both structures using ZF precoding considering quantized analog and digital precoding are characterized. Most existing literature focuses on the fully-connected structure because it is believed to enjoy better performance compared to the sub-connected array \cite{7248552}, while the dissipation due to the use of power dividers/combiners was rarely considered. There is no doubt that dissipation is inevitable during the signal analysis if accurate performance analysis is desired, especially for microwave and mmWave systems. We discover that, under the consideration of dissipation, the sub-connected structure is able to reduce hardware complexity, surprisingly, with improved performance in comparison with the fully-connected structure using a massive antenna array.

2) We employ a channel correlation-based codebook for the hybrid processing system. The advantages of channel correlation-based codebook have already been highlighted in multiuser MIMO \cite{1715541}\cite{8008852, 4698510, 5458368, 1608648}. However, the channel correlation-based codebook has more advantages in hybrid precoding than in conventional multiuser MIMO systems. In hybrid precoding, it is revealed that the effective channels will be correlated even though the physical channels are uncorrelated. 
Therefore, the conventional RVQ codebook is not suitable for hybrid processing. To help design the quantization codebook for the digital precoding, we derive the closed-form expression of the correlation matrix of the effective channels incurred by analog precoding. Moreover, by deriving the spectral efficiency of the considered hybrid precoding system, the channel correlation-based codebook is proved to save feedback bits in comparison with the RVQ codebook in hybrid processing as validated by both theoretical analysis and simulation results. Particularly, the difference between the number of feedback bits needed by the channel correlation-based codebook and that of the RVQ increases with the number of antennas.

3) The effect of factors, including feedback bits, signal power and the number of antennas, on system performance is analyzed. We reveal that the performance loss caused by quantized analog precoding can be compensated by increasing signal power. Moreover, the sub-connected structure 
outperforms the fully-connected structure in either massive MIMO or low signal-to-noise ratio (SNR) scenarios; otherwise, the fully-connected structure achieves better performance.

The rest of the paper is organized as follows. In Section II, system model is described
. In Section III, we analyze the system spectral efficiency performance assuming quantized analog and digital precoding. Section IV compares the sub-connected structure and the fully-connected structure. Simulation results are presented in Section V before concluding 
in Section VI.

Notations throughout this paper: Upper and lower case bold-face letters are matrices and vectors, respectively. Lower case normal letters are scalars. Upper case fraktur letters are sets. $\|\cdot\|$ and $(\cdot)^H$ represent the Frobenius norm and Hermitian of a matrix, respectively. $|\cdot|$ and $(\cdot)^*$ represent the absolute value and conjugation of a complex number, respectively. $\mathbf{I}$ means the identity matrix. $\mathbb{C}^{m\times n}$ and $\mathbb{R}^{m\times n}$ respectively denote the ensemble of complex and real numbers. $\mathbf{1}_T\in\mathbb{R}^{T\times 1}$ and $\mathbf{0}_T\in\mathbb{R}^{T\times 1}$ respectively refer to all-one and all-zero vectors. $\Re[\cdot]$ and $\Im[\cdot]$ represent the real and imaginary parts of a complex number, respectively. $\mathcal{N}(\mu,\sigma^2)$ and $\mathcal{CN}(\mu,\sigma^2)$ respectively stand for the Gaussian distribution and complex Gaussian distribution with mean $\mu$ and variance $\sigma^2$. $\mathbb{E}[\cdot]$, $\mathbb{V}[\cdot]$ and $\mathbb{C}ov[\cdot]$ are respectively used to denote expectation, variance and covariance. $\overset{a.s.}{\rightarrow}$ indicates almost sure convergence. ${\rm diag}(\mathbf{a})$ refers to a diagonal matrix whose diagonals are the entries of vector $\mathbf{a}$.

\section{System Model}

We focus on the downlink of a multiuser massive MIMO using hybrid analog and digital precoding. The system consists of a base station (BS) equipped with $M$ antennas and $K$ RF chains. Different from a typical MIMO setup, 
$K$ is far less than $M$. Note that $K$ is assumed to be greater than or equal to the number of simultaneously served users to enable multi-stream communication. Since the number of overall active users could be in practice arbitrarily large, we assume that the BS first selects $K$ users from the entire active user pool before transmission.

\begin{figure}
\subfigure[]{
\begin{minipage}[b]{0.45\textwidth}
\centering
\includegraphics[width=0.7\textwidth]{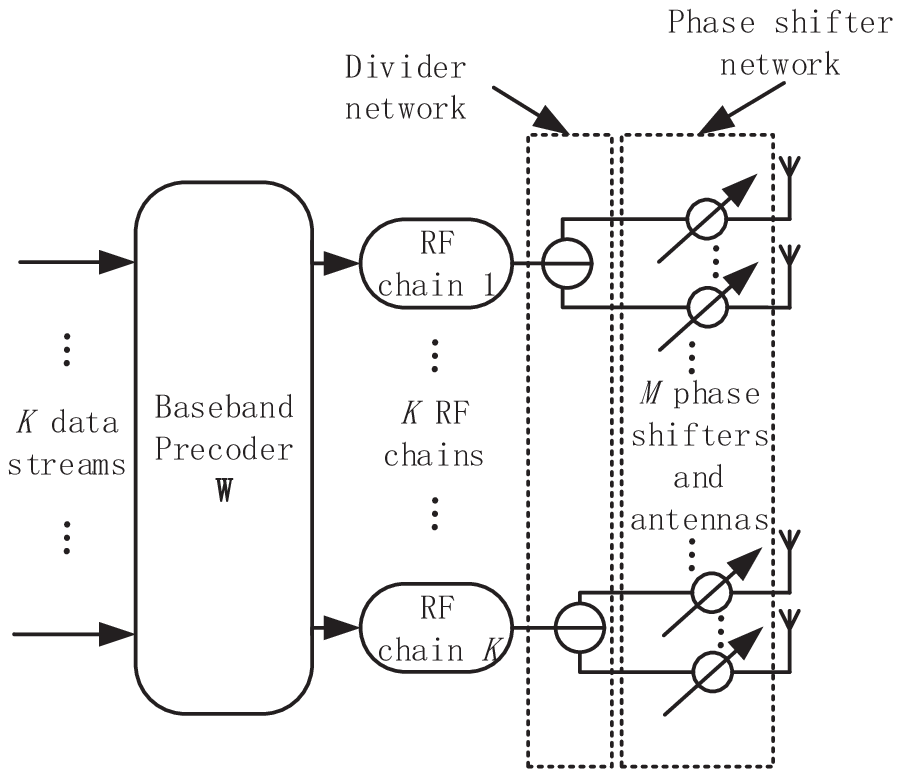}
\end{minipage}
}
\subfigure[]{
\begin{minipage}[b]{0.45\textwidth}
\includegraphics[width=0.9\textwidth]{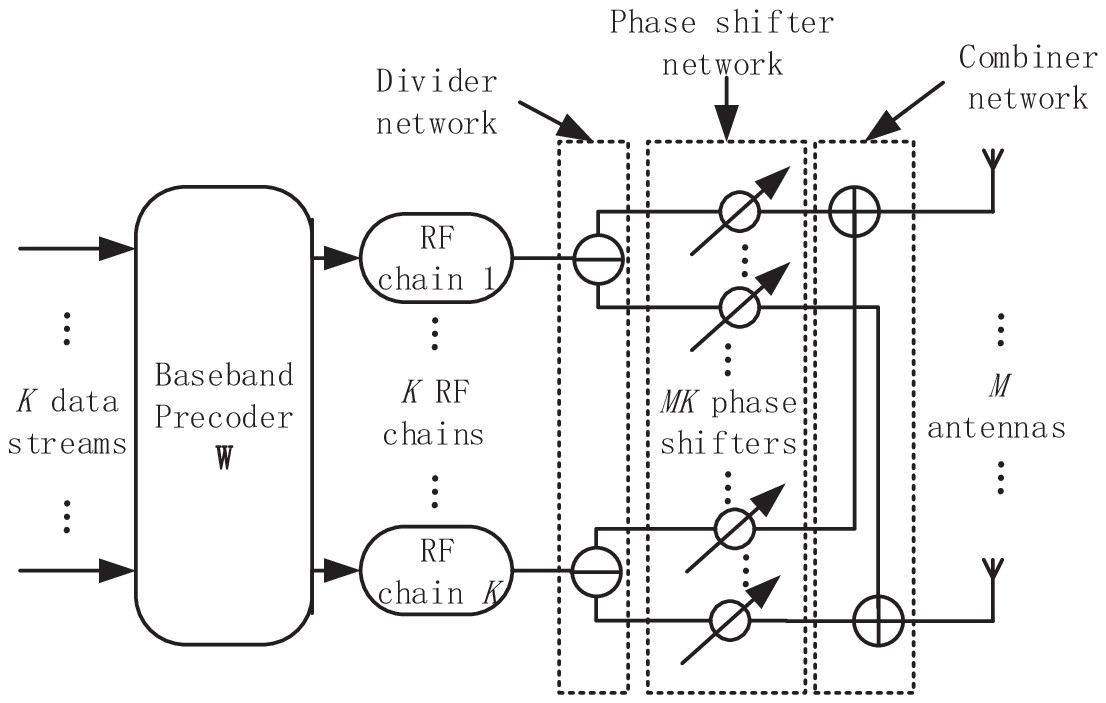}
\end{minipage}
}
 \caption{Hybrid precoding structures for 
massive MIMO systems: (a) the sub-connected structure; (b) the fully-connected structure.}
\end{figure}

At BS, each RF chain is connected to some antennas through phase shifters. There are generally two typical setups of phase shifter networks connecting RF chains with antennas. For the sub-connected structure in Fig. 1(a), each RF chain is connected a disjoint subset of $N=\frac{M}{K}$ antennas and each antenna has its exclusive phase shifter. In contrast, for the fully-connected structure in Fig. 1(b), each RF chain drives all antennas and the signals from all RF chains are combined before being fed to the antenna.

The received signal at all $K$ users can be represented as
\begin{align}
\mathbf{y}=\mathbf{H}^H\mathbf{AWs}+\mathbf{n}\label{eq18}
\end{align}
where $\mathbf{s}\in\mathbb{C}^{K\times 1}$ represents the transmit symbol vector with $\mathbb{E}\{\mathbf{ss}^H\}=\frac{P}{K}\mathbf{I}_K$, in which $P$ is the total initial signal power at BS. $\mathbf{n}$ denotes the additive white Gaussian noise vector with $\mathbf{n}\sim \mathcal{CN}(\mathbf{0}_K,\mathbf{I}_K)$. $\mathbf{H}=[\mathbf{h}_1,\mathbf{h}_2,...,\mathbf{h}_K]\in\mathbb{C}^{M\times K}$ denotes the channel matrix between BS and all users with $\mathbf{h}_k\sim\mathcal{CN}(\mathbf{0}_M,\mathbf{I}_M)$, $\mathbf{A}=[\mathbf{a}_1,\mathbf{a}_2,...,\mathbf{a}_K]\in\mathbb{C}^{M\times K}$ represents the analog precoding network matrix, and $\mathbf{W}=[\mathbf{w}_1,\mathbf{w}_2,...,\mathbf{w}_K]\in\mathbb{C}^{K\times K}$ is the digital precoding matrix.

\subsection{Analog Precoding Network with Dissipation}

As depicted in Fig. 1, the signal from each RF chain is first divided into multiple equal-power outputs to phase shifters. In the sub-connected structure, the phase shifted signals are directly transmitted, whereas in the fully-connected structure, the signals are combined before transmission. It suggests that the entire analog precoding network is composed of divider network, phase shifter network and combiner network in the fully-connected structure. In contrast, the analog precoding network is similar in the sub-connected structure except that there is no combiner network.

In our work, we consider the dissipation to characterize the power changes through power dividers and combiners. For the purpose of characterizing the changes of signals through hardware circuits, it is necessary to separately describe the divider, phase shifter and combiner networks based on the S-parameter of the hardware components.

In the sub-connected structure, the effect of analog processing can be characterized by \cite{7482822}
\begin{align}
\mathbf{A}&=\mathbf{F}_{PS}\mathbf{F}_D\label{eq01}
\end{align}
where $\mathbf{F}_{PS}\in\mathbb{C}^{M\times M}$ denotes the matrix operation by the phase shifter network and $\mathbf{F}_D\in\mathbb{R}^{M\times K}$ stands for the effect of the divider network. Assuming that popular Wilkinson power dividers and combiners are utilized, the reflected power from the output ports is dissipated. Accordingly, the impact of the divider network, relating to the number of ports \cite{001}, 
can be expressed by
\begin{align}
\mathbf{F}_D=\sqrt{\frac{1}{N}}\left[
\begin{matrix}
\mathbf{1}_N&\mathbf{0}_N&\ldots&\mathbf{0}_N\\
\mathbf{0}_N&\mathbf{1}_N&\ldots&\mathbf{0}_N\\
\vdots&\vdots&\ddots&\vdots\\
\mathbf{0}_N&\mathbf{0}_N&\ldots&\mathbf{1}_N
\end{matrix}
\right].\label{eq16}
\end{align}
Similarly, the precoding matrix of the phase shifter network can be given by
\begin{align}
\mathbf{F}_{PS}=\left[
\begin{matrix}
{\rm diag}(\mathbf{f}_{1}^{PS})&{\rm diag}(\mathbf{0}_N)&\ldots&{\rm diag}(\mathbf{0}_N)\\
{\rm diag}(\mathbf{0}_N)&{\rm diag}(\mathbf{f}_{2}^{PS})&\ldots&{\rm diag}(\mathbf{0}_N)\\
\vdots&\vdots&\ddots&\vdots\\
{\rm diag}(\mathbf{0}_N)&{\rm diag}(\mathbf{0}_N)&\ldots&{\rm diag}(\mathbf{f}_{K}^{PS})
\end{matrix}
\right]\label{eq17}
\end{align}
where $\mathbf{f}_{k}^{PS}\in\mathbb{C}^{N\times1}$ represents the function of phase shifters connected to the $k$-th RF chain.

Substituting \eqref{eq16} and \eqref{eq17} into \eqref{eq01}, we now have the analog processing
in the sub-connected structure as
\begin{align}
\mathbf{A}=\sqrt{\frac{1}{N}}\left[
\begin{matrix}
\mathbf{f}_{1}^{PS}&\mathbf{0}_N&\ldots&\mathbf{0}_N\\
\mathbf{0}_N&\mathbf{f}_{2}^{PS}&\ldots&\mathbf{0}_N\\
\vdots&\vdots&\ddots&\vdots\\
\mathbf{0}_N&\mathbf{0}_N&\ldots&\mathbf{f}_{K}^{PS}
\end{matrix}
\right]\overset{\triangle}{=}\sqrt{\frac{1}{N}}\mathbf{F}_{sub}\label{eq03}
\end{align}
where $\mathbf{F}_{sub}$ stands for the equivalent analog precoding matrix in the sub-connected structure which is implemented by phase shifters.

On the other hand, in the fully-connected structure, the effect of combiner network should also be considered. Similarly, the analog processing $\mathbf{A}$ takes the form
\begin{align}
\mathbf{A}=\mathbf{F}_C\mathbf{F}_{PS}\mathbf{F}_D=\sqrt{\frac{1}{MK}}[\mathbf{f}_{1}^{PS},\mathbf{f}_{2}^{PS},\ldots,\mathbf{f}_{K}^{PS}]\overset{\triangle}{=}\sqrt{\frac{1}{MK}}\mathbf{F}_{full}\label{eq06}
\end{align}
where $\mathbf{F}_C=\sqrt{\frac{1}{K}}[{\rm diag}(\mathbf{1}_M), {\rm diag}(\mathbf{1}_M), ..., {\rm diag}(\mathbf{1}_M)]$ denotes the matrix operation by the combiner network, $\mathbf{F}_{full}$ stands for the phase shifter matrix in the fully-connected structure and $\mathbf{f}_{k}^{PS}\in\mathbb{C}^{M\times1}$ has the same definition as in \eqref{eq03}, but a different dimension from that in 
\eqref{eq03}.

\subsection{Quantized Hybrid Precoding Design}

Speaking of designing analog precoding using phase shifters, the angle elements are usually quantized and selected from a finite-size codebook because most phase shifters are digitally-controlled in current communication systems \cite{6717211, 7160780, 7510972}. In this paper, the angle of each phase shifter is chosen from a codebook $\mathcal{A}=\{e^{j2\pi n/2^{B_1}},n=0,1,...,2^{B_1}-1\}$ based on the minimum Euclidean distance criterion where $B_1$ represents the number of quantization bits for analog precoding per phase shifter.

Generally, for the phase shifter network in the sub-connected structure, the $j$-th element of $\hat{\mathbf{f}}_k^{PS}$, i.e., quantized version of $\mathbf{f}_k^{PS}$, is normalized by
\begin{align}
\hat{f}_{k,j}^{PS}=\frac{1}{\sqrt{N}}e^{j\hat{\varphi}_{k,j}}\label{eq07}
\end{align}
where $\hat{\varphi}_{k,j}$ is a quantized angle belonging to $\mathcal{A}$. Combining \eqref{eq03} and \eqref{eq07}, the ($i,j$)-th element of $\hat{\mathbf{A}}$, i.e., quantized version of $\mathbf{A}$, can be written by
\begin{align}
\hat{a}_{k,j}=\frac{1}{\sqrt{N}}\hat{f}_{k,j}=\left\{\begin{array}{ll}
\frac{1}{N}e^{j\hat{\varphi}_{k,j}}, & N(k-1)+1\leq j<Nk \\
0, & {\rm otherwise}
\end{array}\right.\label{eq02}
\end{align}
where $\hat{f}_{k,j}$ refers to the ($i,j$)-th element of $\hat{\mathbf{F}}$, i.e., quantized version of $\mathbf{F}$.

The analog precoding is designed by selecting optimal angles to maximize the signal power of each user. In this stage, the analog precoding enlarges the received signal power as much as possible and the interference among users is left to 
the following digital precoding. Therefore, for the sub-connected structure, the analog precoding is implemented by selecting $\hat{a}_{k,j}$ via
\begin{align}
\hat{a}_{k,j}=\left\{\begin{array}{ll}
\frac{1}{N}\mathop{\argmax}\limits_{e^{j\hat{\varphi}_{k,j}}\in\mathcal{A}}{|h_{k,j}^*e^{j\hat{\varphi}_{k,j}}|}, & N(k-1)+1\leq j<Nk \\
0, & {\rm otherwise}
\end{array}\right.\label{eq19}
\end{align}
where $h_{k,j}$ is the $j$-th element of $\mathbf{h}_{k}$. It is notable that the normalization of analog precoding is only imposed on the elements of $\mathbf{F}$ but not the entries of $\mathbf{A}$. That is to conduct normalization on analog precoding but not representing the impacts of dividers and combiners. Therefore, the normalization does not cover the effects of the signal changes (dissipation) due to power dividers and combiners.

In contrast, for the phase shifter network in the fully-connected structure, $\hat{a}_{k,j}$ is given by
\begin{align}
\hat{a}_{k,j}=\frac{1}{M\sqrt{K}}e^{j\hat{\varphi}_{k,j}}=\frac{1}{M\sqrt{K}}\mathop{\argmax}\limits_{e^{j\hat{\varphi}_{k,j}}\in\mathcal{A}}{|h_{k,j}^*e^{j\hat{\varphi}_{k,j}}|}.\label{eq32}
\end{align}

The digital precoding follows the ZF criterion which cancels interference among users. Note that the ZF precoding for both structures shares the same form and is dependent on the effective channels. For user $k$, its effective channel is defined as
\begin{align}
\mathbf{g}_k^H\overset{\triangle}{=}\mathbf{h}_k^H\hat{\mathbf{A}}.\label{eq22}
\end{align}
Each user $k$ quantizes its effective channel vector using a codebook, $\mathcal{G}$, of size $2^{B_2}$ according to $\hat{\mathbf{g}}_k=\mathop{\argmax}_{\hat{\mathbf{g}}_k\in \mathcal{G}}{|\mathbf{g}_k^H\hat{\mathbf{g}}_k|}$ which gives a quantized version of $\mathbf{g}_k$. Each user feeds back the quantized effective channel vector, $\hat{\mathbf{g}}_k$, with $B_2$ bits. According to the feedback from all users, the BS calculates the unnormalized ZF precoder as
\begin{align}
\hat{\mathbf{U}}=\hat{\mathbf{G}}(\hat{\mathbf{G}}^H\hat{\mathbf{G}})^{-1}\label{eq20}
\end{align}
where $\hat{\mathbf{G}}=[\hat{\mathbf{g}}_1,\hat{\mathbf{g}}_2,...,\hat{\mathbf{g}}_K]$ and $\hat{\mathbf{U}}=[\hat{\mathbf{u}}_1,\hat{\mathbf{u}_2},...,\hat{\mathbf{u}}_K]$. The digital precoder is finally normalized as $\hat{\mathbf{w}}_k=\frac{\hat{\mathbf{u}}_k}{\|\hat{\mathbf{F}}\hat{\mathbf{u}}_k\|}$  to fulfill the power constraints, i.e., $\|\hat{\mathbf{F}}\hat{\mathbf{w}}_k\|=1$, $k=1,2,...,K$, where $\hat{\mathbf{W}}=[\hat{\mathbf{w}}_1,\hat{\mathbf{w}_2},...,\hat{\mathbf{w}}_K]$. Note that $\mathbf{W}$ and $\hat{\mathbf{W}}$ respectively denote digital precoding matrices based on perfect effective channel feedback and quantized effective channel feedback.

\section{Spectral Efficiency Performance Analysis with Quantized Precoding}

\subsection{Quantized Analog Precoding with Full Channel State Information (CSI) Feedback}
Since the hybrid precoding consists of the analog precoding and digital precoding, it is reasonable to discuss the effect of quantization for the analog precoding and digital precoding separately. In this subsection, we analyze the quantized hybrid precoding with quantized analog precoding and unquantized digital precoding.

Assuming perfect digital precoding quantization and considering the nature of ZF, no multiuser interference 
exists. From \eqref{eq18}, \eqref{eq19}, \eqref{eq32} and \eqref{eq20}, the received signal at the $k$-th user is
\begin{align}
y_k=\mathbf{h}_k^H\hat{\mathbf{A}}\mathbf{w}_ks_k+n_k
\end{align}
where $\mathbf{w}_k$ denotes the $k$-th column of $\mathbf{W}$.

Before analyzing the spectral efficiencies, we first present Lemma 1 about the effective channels since the digital precoding is mainly relevant to them.

\begin{myLem}
Let $\mathbf{G}=[\mathbf{g}_1,\mathbf{g}_2,...,\mathbf{g}_k]$ be the effective channel matrix. For massive MIMO with growing numbers of antennas to infinity, we have
\begin{align}
\mathbf{G}\overset{a.s.}{\rightarrow}\rho\mathbf{I},\rho\in\{\rho^{sub},\rho^{full}\}\label{eq23}
\end{align}
in which $\rho^{sub}=\frac{\sqrt{\pi}}{2}{\rm sinc}\left(\frac{\pi}{2^{B_1}}\right)$ is for the sub-connected structure and $\rho^{full}=\sqrt{\frac{\pi}{4K}}{\rm sinc}\left(\frac{\pi}{2^{B_1}}\right)$ is for the fully-connected structure.
\end{myLem}
\emph{Proof:} See Appendix A.

Then, the asymptotic spectral efficiency of the $k$-th user can be obtained by
\begin{align}
R_k&=\log_2\left(1+\frac{P}{K}|\mathbf{h}_k^H\hat{\mathbf{A}}\mathbf{w}_k|^2\right)\nonumber\\
&=\log_2\left(1+\frac{P}{K}|\mathbf{g}_k^H\mathbf{w}_k|^2\right)\nonumber\\
&\overset{(a)}{\overset{a.s.}{\rightarrow}}\log_2\left(1+\frac{P}{K}|g_{k,k}^*|^2\right)\nonumber\\
&\overset{(b)}{=}\log_2\left(1+\frac{P}{K}\rho^2\right)\label{eq21}
\end{align}
where (a) follows from $\mathbf{W}\overset{a.s.}{\rightarrow}\mathbf{I}$ in which \eqref{eq20} and \eqref{eq23} are utilized and (b) uses \eqref{eq23}. As we normalize the noise power as $1$, $P$ in \eqref{eq21} corresponds to the SNR where the power of noise is included. Note that the expression in \eqref{eq21} is obtained as the asymptotic behavior of hybrid precoding performance with respect to the antenna number. The larger the number of antennas at BS is, the more accurate \eqref{eq21} becomes. Note that this observation, in some sense, coincides with the fundamental knowledge on massive MIMO that multiuser channels are asymptotically orthogonal for infinite $M$ but not finite $M$ \cite{5595728}.

Consequently, we can conclude in the following Theorem on the spectral efficiencies of hybrid precoding with only quantized analog precoding from \eqref{eq21}.
\begin{myTheo}
For the massive MIMO system, the spectral efficiencies of hybrid precoding under the sub-connected and the fully-connected structures can be respectively characterized by
\begin{align}
R_k^{sub}&\overset{a.s.}{\rightarrow}\log_2\left(1+\frac{\pi P}{4K}{\rm sinc}^2\left(\frac{\pi}{2^{B_1}}\right)\right),\ R_k^{full}\overset{a.s.}{\rightarrow}\log_2\left(1+\frac{\pi P}{4K^2}{\rm sinc}^2\left(\frac{\pi}{2^{B_1}}\right)\right).\label{eq25}
\end{align}
\end{myTheo}

\begin{myRem}
The fully-connected structure has in general been regarded as enjoying better performance because a fully-connected phase shifter network can realize more accurate analog beamforming. As discovered in Theorem 1, higher dissipation in the fully-connected structure caused by the divider network, as signals being divided into more streams, however, cancels out the benefits in SINR acquired by accurate analog beamforming. Moreover, the extra combiner network in the fully-connected structure leads to further deterioration in the received SINR. \emph{It is shown in Theorem 1 that the SINR for the sub-connected structure is $K$ times as large as that for the fully-connected structure by taking the practical dissipation into account. Therefore, the sub-connected structure surprisingly achieves a higher spectral efficiency due to the additional dissipation caused by RF circuits, especially under the scenario with a massive antenna array, compared to the fully-connected structure.}
\end{myRem}

\subsection{Some Insights Regarding $K$}

As we have derived some expressions on the spectral efficiency, we can observe some insights on the number of RF chains. We first investigate the sum spectral efficiency of the sub-connected structure. According to the analysis in the last subsection, the sum spectral efficiency with perfect digital quantization is represented as
\begin{align}
f(K)\overset{\triangle}{=}KR_k^{sub}.
\end{align}
For the sake of analysis, we temporarily relax $K$ as a continuous positive variable. Then we are able to check the derivative of $f(K)$ with respect to $K$. It gives $f'(K)=\log_2(1+\frac{\xi}{K})-\frac{\xi}{(K+\xi)\ln2}$ and
$f''(K)=\frac{\xi}{(K+\xi)\ln2}\left(\frac{1}{K+\xi}-\frac{1}{K}\right)$ where $\xi=\frac{\pi P}{4}{\rm sinc}^2(\frac{\pi}{2^{B_1}})$. It is not hard to check
\begin{align}
f''(K)<0,\ \forall K>1.\label{04}
\end{align}

From \eqref{04}, we know that $f'(K)$ is a monotonic decreasing function for $K>1$. In order to determine the range of $f'(K)$, we first need the following result:
\begin{align}
\lim\limits_{K\to+\infty}f'(K)=&\lim\limits_{K\to+\infty}\log_2(1+\frac{\xi}{K})-\frac{\xi}{(K+\xi)\ln2}\nonumber\\
=&\frac{1}{\ln2}\lim\limits_{K\to+\infty}\left[\ln(1+\frac{\xi}{K})-\frac{\xi}{K+\xi}\right]\nonumber\\
\overset{(a)}{\rightarrow}&\frac{1}{\ln2}\lim\limits_{K\to+\infty}\left[\frac{\xi}{K}+o(\frac{\xi}{K})-\frac{\xi}{K+\xi}\right]\nonumber\\
\approx&\frac{1}{\ln2}\lim\limits_{K\to+\infty}\left[\frac{\xi}{K}-\frac{\xi}{K+\xi}\right]\nonumber\\
>&0\label{06}
\end{align}
where (a) utilizes the Taylor's expansion for $\ln\left(1+\frac{\xi}{K}\right)$. Combining \eqref{04} and \eqref{06}, we know that $f'(K)>0$ is always true when $K>1$, which indicates that $f(K)$ is a monotonic increasing function for $K>1$. That is to say, the sum spectral efficiency monotonically increases with $K$.

Similarly as derived for the sub-connected structure, we reveal that, for the fully-connected structure, the sum spectral efficiency decreases with $K$ for $\xi\leq3.92K^2$, while for $\xi>3.92K^2$, the sum spectral efficiency increases with $K$.

\subsection{Digital Precoding with Quantized Feedback}

In many researches on TDD-based massive MIMO, the downlink CSI can be directly obtained by applying channel reciprocity. However, uplink and downlink channels, in practice, may not be reciprocal due to 
frequency response mismatches between the transmitter and receiver chains. Unavoidable physical limitations of the used electronics \cite{razavi1998rf} make the channel reciprocity difficult to be achieved, especially if low-cost transceivers are expected to be deployed and implemented at the BS to keep the total cost feasible \cite{6736761} \cite{6736746} \cite{6375940}. Therefore, we assume that the massive MIMO system considered in this work could not hold the reciprocity due to the hardware impairments which implies developing quantized digital precoding relies on quantized CSI feedback. In massive MIMO, although channel estimation could be challenging in practice, there have been some schemes \cite{6777295} \cite{6816089} \cite{7174558}, especially for hybrid architecture based massive MIMO with reduced overhead \cite{7321002}. Concerning the pilot quantity for massive MIMO, the overhead of pilots roughly amounts to $\mathcal{O}(M)$. We make the assumption that the system runs in a low mobility scenario where the channel block length is large.

It is notable that, in hybrid processing, digital precoding is designed based on the effective channels, defined in \eqref{eq22}. Most existing research focuses on the RVQ codebook in MIMO systems under the assumption that channels are isotropically distributed. However, the assumption does not apply to the effective channels in hybrid processing. Although RVQ simplifies the analysis process and allows leveraging some results from the limited feedback MIMO literature
\cite{1715541} \cite{4641946} \cite{5504192}, channel correlation-based codebook is expected to be preferrable in hybrid processing because the channel correlation-based codebook is able to characterize channel correlation information which results from the fact that the effective channels are always affected by the analog precoding. Even though the physical channels could be ideally Rayleigh fading, the effective channel, however, can no longer be isotropic after being processed by the analog precoding network. The channel correlation-based codebook in a multi-antenna setup was first introduced and analyzed in \cite{1715541}. In \cite{8008852}, the channel correlation-based codebooks have been studied and used in hybrid precoding. However, the correlation matrix is not explicitly analyzed, neither is its impact on the spectral efficiency loss resulting from the feedback which will be analyzed in 
this section. The structure of the codebooks utilized in this paper, which was first developed for conventional multiuser MIMO systems in \cite{1608648}, will be introduced in the following.

Define the correlation matrix of the $k$-th user's effective channel as $\mathbf{R}_{k}\in \mathbb{C}^{K\times K}$. The $k$-th user's quantization vector $\mathbf{c}_{k,i}$ can be obtained by multiplying an independent and identically distributed (i.i.d.) complex Gaussian vector $\mathbf{v}_i\in \mathbb{C}^{K\times 1}, i\in \{1, 2,..., 2^{B_2}\}$ chosen from an RVQ codebook by the square root of the channel correlation matrix i.e., $\mathbf{R}_{k}^{1/2}\mathbf{v}_i$. The quantization vector is further normalized as
\begin{align}
\mathbf{c}_{k,i}=\frac{\mathbf{R}_{k}^{1/2}\mathbf{v}_i}{\|\mathbf{R}_{k}^{1/2}\mathbf{v}_i\|}.
\end{align}

As $\mathbf{R}_{k}$ is the correlation matrix of the $k$-th user's effective channel, we obtain
$\mathbf{R}_{k}=\mathbb{E}[\mathbf{g}_k\mathbf{g}_k^H]$
which further yields
\begin{align}
r_{i,j}=\mathbb{E}[g_{k,i}g_{k,j}^*]\label{eq27}
\end{align}
where $r_{i,j}$ ($i,j=1,2,...,K$) is the $(i,j)$-th element of $\mathbf{R}_{k}$. Obviously, the key point of the channel correlation-based codebook is to explicitly evaluate the correlation matrix of the $k$-th user's effective channel.

\begin{myLem}
The asymptotic correlation matrix of the $k$-th user's effective channel can be represented as $\mathbf{R}_{k}={\rm diag}([r_{1,1},r_{2,2},...,r_{K,K}]^T)$. For the sub-connected structure,
\begin{align}
r_{i,i}&=\left\{\begin{array}{ll}
\frac{\pi}{4}{\rm sinc}^2\left(\frac{\pi}{2^{B_1}}\right)+\frac{1}{N}-\frac{\frac{\pi}{4}{\rm sinc}^2\left(\frac{\pi}{2^{B_1}}\right)}{N}, & i=k \\
\frac{1}{N}, & {\rm otherwise},
\end{array}\right.\label{eq12}
\end{align}
whereas for the fully-connected structure,
\begin{align}
r_{i,i}&=\left\{\begin{array}{ll}
\frac{\pi}{4K}{\rm sinc}^2\left(\frac{\pi}{2^{B_1}}\right)+\frac{1}{MK}-\frac{\frac{\pi}{4}{\rm sinc}^2\left(\frac{\pi}{2^{B_1}}\right)}{MK}, & i=k \\
\frac{1}{MK}, & {\rm otherwise}.
\end{array}\right.\label{eq09}
\end{align}
\end{myLem}
\emph{Proof:} See Appendix B.

Because of the appearance of the channel correlation-based codebook, it becomes difficult to directly characterize the spectral efficiency performance under various structures. Alternatively, we investigate the behavior of the system spectral efficiency via characterizing the spectral efficiency loss due to the finite-rate feedback of effective channels. Note that, mathematically, we define the spectral efficiency loss as the difference between the system spectral efficiency achieved with only quantized analog precoding and the one achieved with both quantized analog precoding and digital precoding using the channel correlation-based codebook. By exploiting the new results in Lemma 2, we have following results on the spectral efficiency loss bounds.

\begin{myTheo}
For massive MIMO systems adopting hybrid precoding and $B_2$ feedback bits per user, the average spectral efficiency losses of each user in the sub-connected and the fully-connected structures are respectively upper bounded by
\begin{align}
\Delta R_{sub}\leq\log_2\left(1+\frac{P(K-1)}{M}2^{-\frac{B_2}{K-1}}\right),\ \Delta R_{full}\leq\log_2\left(1+\frac{P(K-1)}{MK^2}2^{-\frac{B_2}{K-1}}\right).\label{eq28}
\end{align}
\end{myTheo}
\emph{Proof:} See Appendix C.

It is interesting to find that $B_1$, which is related to the channel correlation information, does not explicitly affect the spectral efficiency loss, i.e. $\Delta R_{sub}$ and $\Delta R_{full}$. In this work, the physical channels are uncorrelated and the correlation of effective channels is caused by the phase shifter network. From \eqref{eq28}, if the quantized codebook is designed according to the correlation of effective channels, the impact of the phase shifter network will be eliminated by the effect of quantized digital precoding. Then regarding the effect of signal power with dissipation, as explained in Remark 1, the dissipation in the sub-connected structure is smaller than that in the fully-connected structure
. Therefore, the signal power is larger in the sub-connected structure which results in a higher spectral efficiency loss.

From (16), the sub-connected structure outperforms in spectral efficiency with ideal (unquantized) digital precoding ($R_k^{sub}>R_k^{full}$). While when quantized digital precoding is considered, the fully-connected structure always has less spectral efficiency loss ($\Delta R_{sub}<\Delta R_{full}$) compared to the loss of the sub-connected structure. Thus, it is difficult to simply point out which one is better, which will be discussed in the next section. In particular, we additionally presented the potential insights as follows:

1) As explained in Remark 1, the signal power with dissipation in the fully-connected structure is much smaller than in the sub-connected structure. It results in that the spectral efficiency loss in the fully-connected structure changes less remarkably with respect to $B_2$. It implies that, for the fully-connected structure, the spectral loss is affected less significantly than that in the sub-connected structure with respect to a decreasing $B_2$. For the ease of explaining the relationship between $R_k^{full}$ and $\Delta R_{full}$, we give an example with system parameters $M=64$, $P=25$dB, $B_1=3$ and $K=4$. It yields $R_k^{full}\approx3.98$ bits/Hz. Based on (24), we obtain $\Delta R_{full}\approx0.37$ bits/Hz for $B_2=5$ and $\Delta R_{full}\approx0.13$ bits/Hz for $B_2=10$. The performance difference is only $6\%$ since $\frac{\Delta R_{full}}{R_k^{full}}\approx0.093$ for $B_2=5$ and $\frac{\Delta R_{full}}{R_k^{full}}\approx0.033$ for $B_2=10$.

2) For the sub-connected structure, the influence of $B_2$ on the spectral efficiency loss is much more significant than in the fully-connected structure. With the same setup in the above example, we have $R_k^{sub}\approx5.91$ bits/Hz, $\Delta R_{sub}\approx2.50$ bits/Hz for $B_2=5$ and $\Delta R_{sub}\approx1.30$ bits/Hz for $B_2=10$. $\frac{\Delta R_{sub}}{R_k^{sub}}$ decreases from $0.42$ ($B_2=5$) to $0.21$ ($B_2=10$) where difference ($21\%$) is much larger than that in the fully-connected structure. Therefore, the resolution of quantized digital precoding ($B_2$) in the sub-connected structure is suggested to be not too small since it affects the spectral efficiency more remarkably than in the fully-connected structure.
We compare the above two examples and discover that, in applications, we may prefer to improve the resolution of the quantized digital precoding ($B_2$) for the sub-connected structure.

\subsection{Effect of Quantization Bits}

Since we have derived the spectral efficiencies with unquantized digital precoding and the upper bounds of spectral efficiency losses due to quantized digital precoding, we can gain a deeper understanding of the required number of quantization bits for different structures. The average spectral efficiencies achieved with both quantized analog and digital precoding under the sub-connected and fully-connected structures are respectively expressed by
\begin{align}
\bar{R}_{sub}^Q=&R_k^{sub}-\Delta R_{sub},\
\bar{R}_{full}^Q=R_k^{full}-\Delta R_{full}.\label{eq29}
\end{align}

In terms of the effect of quantization bits, from \eqref{eq25} and \eqref{eq28}, we find that $R_k^{sub}$ and $R_k^{full}$ are only related towards $B_1$, and $\Delta R_{sub}$ and $\Delta R_{full}$ are only associated with $B_2$. Therefore, the analysis can be simplified by investigating two independent parts. We first discuss the effect of $B_1$ on $R_k^{sub}$ and $R_k^{full}$, and then the impact of $B_2$ on $\Delta R_{sub}$ and $\Delta R_{full}$ in this subsection.

To maintain $R_k^{sub}$ or $R_k^{full}$ as $\log_2b_1$ bps/Hz per user, from \eqref{eq25}, it holds true that
\begin{align}
{\rm sinc}^2\left(\frac{\pi}{2^{B_1}}\right)=\frac{4K^{1+\zeta}}{\pi P}(b_1-1)
\end{align}
in which $\zeta=0$ for the sub-connected structure and $\zeta=1$ for the fully-connected structure.
By applying the Taylor's expansion to ${\rm sin}^2\left(\frac{\pi}{2^{B_1}}\right)$, we acquire ${\rm sin}^2\left(\frac{\pi}{2^{B_1}}\right)\approx\left(\frac{\pi}{2^{B_1}}\right)^2-\frac{\left(\frac{\pi}{2^{B_1}}\right)^4}{3}$ which implies ${\rm sinc}^2\left(\frac{\pi}{2^{B_1}}\right)\approx1-\frac{\left(\frac{\pi}{2^{B_1}}\right)^2}{3}$. Then, we further have
\begin{align}
B_1\approx\log_2\frac{\pi}{\sqrt{3}}-\frac{1}{2}\log_2\left[1-\frac{4K^{1+\zeta}}{\pi P}(b_1-1)\right]\label{eq13}
\end{align}
when $b_1\leq\frac{\pi P}{4K^{1+\zeta}}+1$. Some observations can be summarized from \eqref{eq13} in the following remark.

\begin{myRem}
To maintain the desired communications, with a fixed power, the spectral efficiency loss due to the quantized precoding could be compensated by increasing $B_1$. On the other hand, with a fixed $B_1$, the spectral efficiency could also be partially compensated by increasing power. When it comes to the comparison between the two structures, it indicates that \emph{the fully-connected structure requires more analog quantization bits per phase shifter to maintain the same transmission spectral efficiency as the sub-connected structure.}
\end{myRem}

To maintain a spectral efficiency loss of $\log_2b_2$ bps/Hz per user, from \eqref{eq28}, the number of digital quantization bits should satisfy
\begin{align}
\frac{B_2}{K-1}=\frac{\log_210}{10}P_{dB}-\log_2\left(\frac{MK^{2\zeta}}{K-1}\right)-\log_2(b_2-1).\label{eq15}
\end{align}
The following remark is outlined from \eqref{eq15}.

\begin{myRem}
\emph{$B_2$ should increase linearly with $P$ in dB and decrease logarithmically with the number of antennas.} As $\zeta$ is different for the sub-connected structure and the fully-connected structure, it suggests that \emph{fewer quantization bits for digital precoding per user are demanded in the fully-connected structure} in order to maintain the same performance.
\end{myRem}

\subsection{Comparison with RVQ}

In this subsection, we analytically verify that the channel correlation-based codebook outperforms the RVQ codebook in hybrid processing.

For the purpose of comparing two different quantized codebooks, we also investigate the spectral efficiency losses using an RVQ codebook which is given by
\begin{align}
\Delta R_{sub}^{RVQ}\leq\log_2\left(1+\frac{\pi P}{4K}{\rm sinc}^2\left(\frac{\pi}{2^{B_1}}\right)2^{-\frac{B_2}{K-1}}\right),\
\Delta R_{full}^{RVQ}\leq\log_2\left(1+\frac{\pi P}{4K^2}{\rm sinc}^2\left(\frac{\pi}{2^{B_1}}\right)2^{-\frac{B_2}{K-1}}\right)\label{eq31}
\end{align}
where $\Delta R_{sub}^{RVQ}$ and $\Delta R_{full}^{RVQ}$ are the spectral efficiency losses using RVQ in the sub-connected and fully-connected structure, respectively. The proof is given in Appendix D.

Denote $B_2^{RVQ}$ as the required number of digital quantization bits using RVQ if a spectral efficiency loss of $\log_2b_2$ bps/Hz per user is allowed. We can calculate that the difference between the number of bits needed by the RVQ codebook and that needed by the channel correlation-based codebook is represented as
\begin{align}
\frac{\Delta B_2}{K-1}=&\frac{B_2^{RVQ}-B_2}{K-1}=\log_2M+\log_2\frac{\pi{\rm sinc}^2(\frac{\pi}{2^{B_1}})}{4K^{1-\zeta}(K-1)}
\end{align}
which leads to the following remark.

\begin{myRem}
\emph{$\Delta B_2$ increases logarithmically with the number of antennas. That is to say, the more antennas the BS has, the more feedback bits the channel correlation-based codebook saves in comparison with the RVQ codebook.} It suggests that the channel correlation-based codebook is preferable in hybrid processing with a massive antenna array.
\end{myRem}

\section{Sub-Connected or Fully-Connected?}
\subsection{When Should We Use Sub-Connected Structure?}
Since the channel correlation-based codebook is proved to enjoy better performance than the conventional RVQ, it is reasonable to analyze the performance of the corresponding hybrid precoding systems. With the analytical results in the last section, we are able to give some conclusions to guide system designs. In this section, we compare the system performance between the sub-connected and fully-connected structures using the channel correlation-based codebook.

As the two structures have different advantages, it is necessary to discuss whether we should use the sub-connected structure or the fully-connected structure. In order to characterize the difference between the sub-connected and fully-connected structures, we define
\begin{align}
\Delta R_1\overset{\triangle}{=}&\bar{R}_{sub}^Q-\bar{R}_{full}^Q\nonumber\\
\overset{(a)}{=}&(R_k^{sub}-\Delta R_{sub})-(R_k^{full}-\Delta R_{full})\nonumber\\
\overset{(b)}{\approx}&\left(\log_2\left(1+\frac{\pi P}{4K}{\rm sinc}^2\left(\frac{\pi}{2^{B_1}}\right)\right)-\log_2\left(1+\frac{P(K-1)}{M}2^{-\frac{B_2}{K-1}}\right)\right)\nonumber\\
&-\left(\log_2\left(1+\frac{\pi P}{4K^2}{\rm sinc}^2\left(\frac{\pi}{2^{B_1}}\right)\right)-\log_2\left(1+\frac{P(K-1)}{MK^2}2^{-\frac{B_2}{K-1}}\right)\right)\label{eq65}
\end{align}
where (a) is obtained from \eqref{eq29}, and (b) is achieved from \eqref{eq25} and \eqref{eq28}. Letting $\Delta R_1\geq0$ which implies the sub-connected structure is preferred, it yields from \eqref{eq65} that
\begin{align}
KM-(K-1)2^{-\frac{B_2}{K-1}}P\geq\frac{4(K^3-K)2^{-\frac{B_2}{K-1}}}{\pi{\rm sinc}^2\left(\frac{\pi}{2^{B_1}}\right)}.\label{eq33}
\end{align}

Clearly, \eqref{eq33} can be analyzed from two aspects, namely the number of BS antennas and the total initial signal power, i.e., $M$ and $P$, via checking the conditions for $\Delta R_1\geq0$.

\begin{myCas}
If we regard $M$ as the design parameter, it is discovered that we should use the sub-connected, instead of, the fully-connected structure when the number of BS antennas satisfies
\begin{align}
M\geq2^{-\frac{B_2}{K-1}}\left[\frac{4(K^2-1)}{\pi {\rm sinc}^2\left(\frac{\pi}{2^{B_1}}\right)}+\left(1-\frac{1}{K}\right)P\right].
\end{align}
\end{myCas}

It indicates that the number of antennas should increase with $P$ if we use the sub-connected structure. Besides, \emph{when the number of antennas is large enough, the sub-connected structure definitely enjoys better performance which implies that the sub-connected structure in fact suits for massive systems with a large amount of antennas. Otherwise, when the number of antennas is small, the fully-connected structure outperforms the sub-connected structure.}

\begin{myCas}
If we regard $P$ as the design parameter, it gives the condition as
\begin{align}
P\leq M2^{\frac{B_2}{K-1}}\frac{K}{K-1}-\frac{4K(K+1)}{\pi{\rm sinc}^2\left(\frac{\pi}{2^{B_1}}\right)}.
\end{align}
\end{myCas}
\emph{When $P$ is quite small, it is obvious that the sub-connected structure outperforms the fully-connected structure which suggests that the sub-connected structure suits for the energy-saving systems without high signal power.}

Actually, the analog precoding matrix of the sub-connected structure is more sparse than the fully-connected structure which means that the load of analog quantization feedback is heavier for the fully-connected structure. Therefore, in practical applications, the sub-connected structure demands fewer analog quantization bits and could perform even better than expected from the derived expressions.

\subsection{Some Further Discussions}

It is important to note that, in our analysis, we assume that power amplifiers with the same processing gain are utilized for both structures. In applications, the dissipation loss can, to some extent, be compensated by using a power amplifier with a much larger processing gain in the fully-connected structure than the power amplifier used in the sub-connected structure. It is, however, known as a heavy burden on hardware design and cost control. Ideally, if we assume that the available processing gain of power amplifiers can be unlimited, we could try to arrive at a different conclusion on the superiority of the two structures by easily modifying the above derived expressions. For an increasing power amplifier gain in the fully-connected structure, the performance gap between the two structures becomes minor. Assume that the power gain of the fully-connected structure is $\eta$ times as large as that of the sub-connected structure when the two structures enjoy the same performance. By solving $\left(\log_2\left(1+\frac{\pi P}{4K}{\rm sinc}^2\left(\frac{\pi}{2^{B_1}}\right)\right)-\log_2\left(1+\frac{P(K-1)}{M}2^{-\frac{B_2}{K-1}}\right)\right)
-\left(\log_2\left(1+\frac{\pi \eta_ P}{4K^2}{\rm sinc}^2\left(\frac{\pi}{2^{B_1}}\right)\right)-\log_2\left(1+\frac{\eta P(K-1)}{MK^2}2^{-\frac{B_2}{K-1}}\right)\right)=0$, we get $\eta=-\frac{\iota_2}{2\iota_1}+\sqrt{\frac{\iota_3}{\iota_1}+\frac{\iota_2^2}{4\iota_1^2}}$ where $\iota_1=\frac{\pi P^2(K-1)}{4MK^2}{\rm sinc}^2\left(\frac{\pi}{2^{B_1}}\right)2^{-\frac{B_2}{K-1}}$, $\iota_2=\frac{\pi P}{4K^2}{\rm sinc}^2\left(\frac{\pi}{2^{B_1}}\right)+\frac{P(K-1)}{M}2^{-\frac{B_2}{K-1}}$ and $\iota_3=\frac{\pi P}{4K}{\rm sinc}^2\left(\frac{\pi}{2^{B_1}}\right)+\frac{P(K-1)}{MK^2}2^{-\frac{B_2}{K-1}}+\frac{\iota_1}{K}$. It implies that, when the power gain of the fully-connected structure is $\eta$ times as large as that of the sub-connected structure, the performance of both structures tends to the same value for massive MIMO. If the power amplifier gain of the fully-connected structure is more than $\eta$ times that of the sub-connected structure, the fully-connected structure enjoys better spectral efficiency than the sub-connected structure.

\section{Simulation Results}

\subsection{Rayleigh Fading Channels}

In this subsection, we evaluate the performance of quantized hybrid precoding in massive MIMO systems over Rayleigh fading channels.

\begin{figure*}
\begin{minipage}[c]{0.32\textwidth}
\centering
\includegraphics[width=0.9\textwidth]{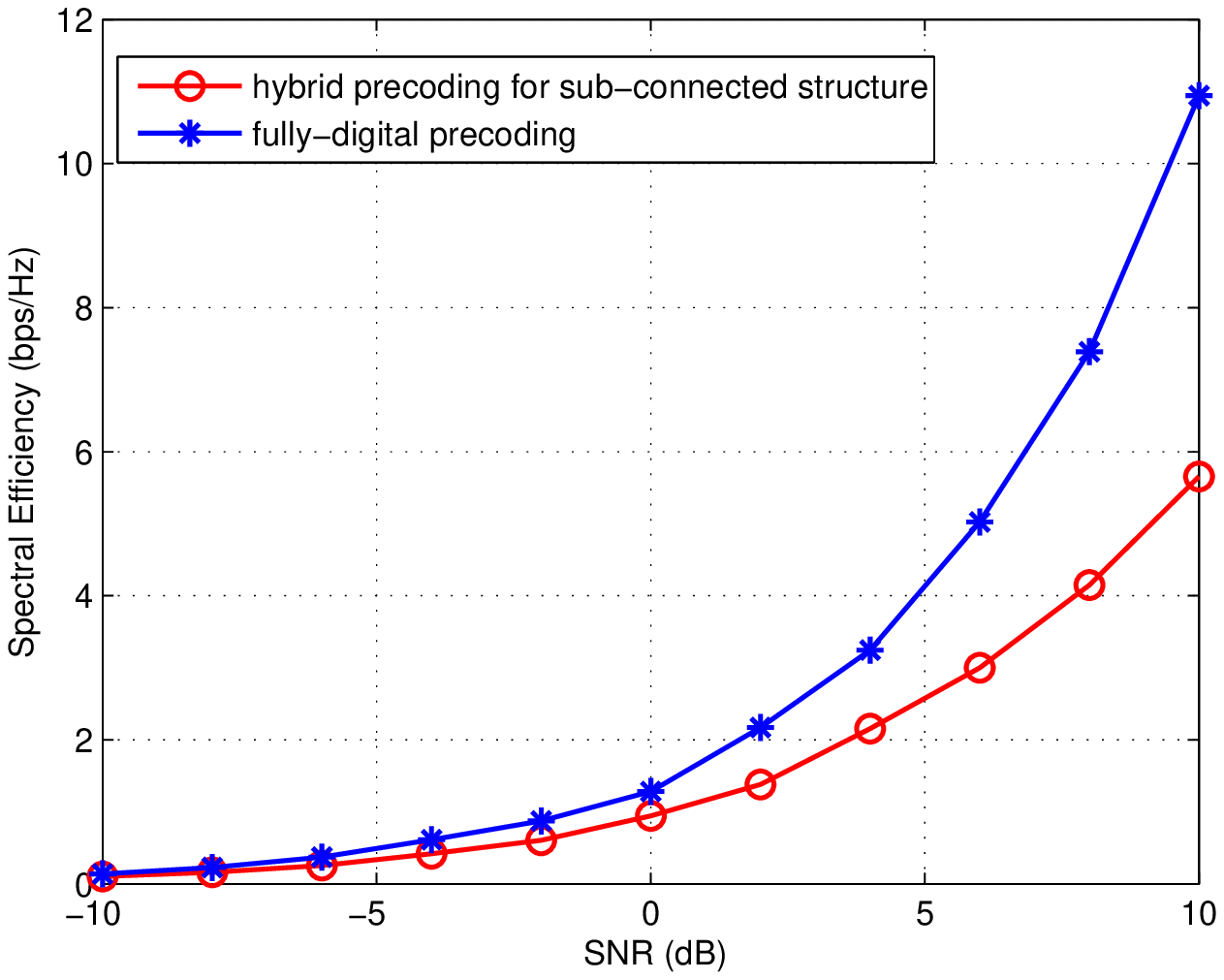}
\caption{Spectral efficiencies using hybrid precoding or fully-digital precoding with $M=64$ and $K=4$.}
\end{minipage}%
\hfill
\begin{minipage}[c]{0.32\textwidth}
\centering
\includegraphics[width=0.9\textwidth]{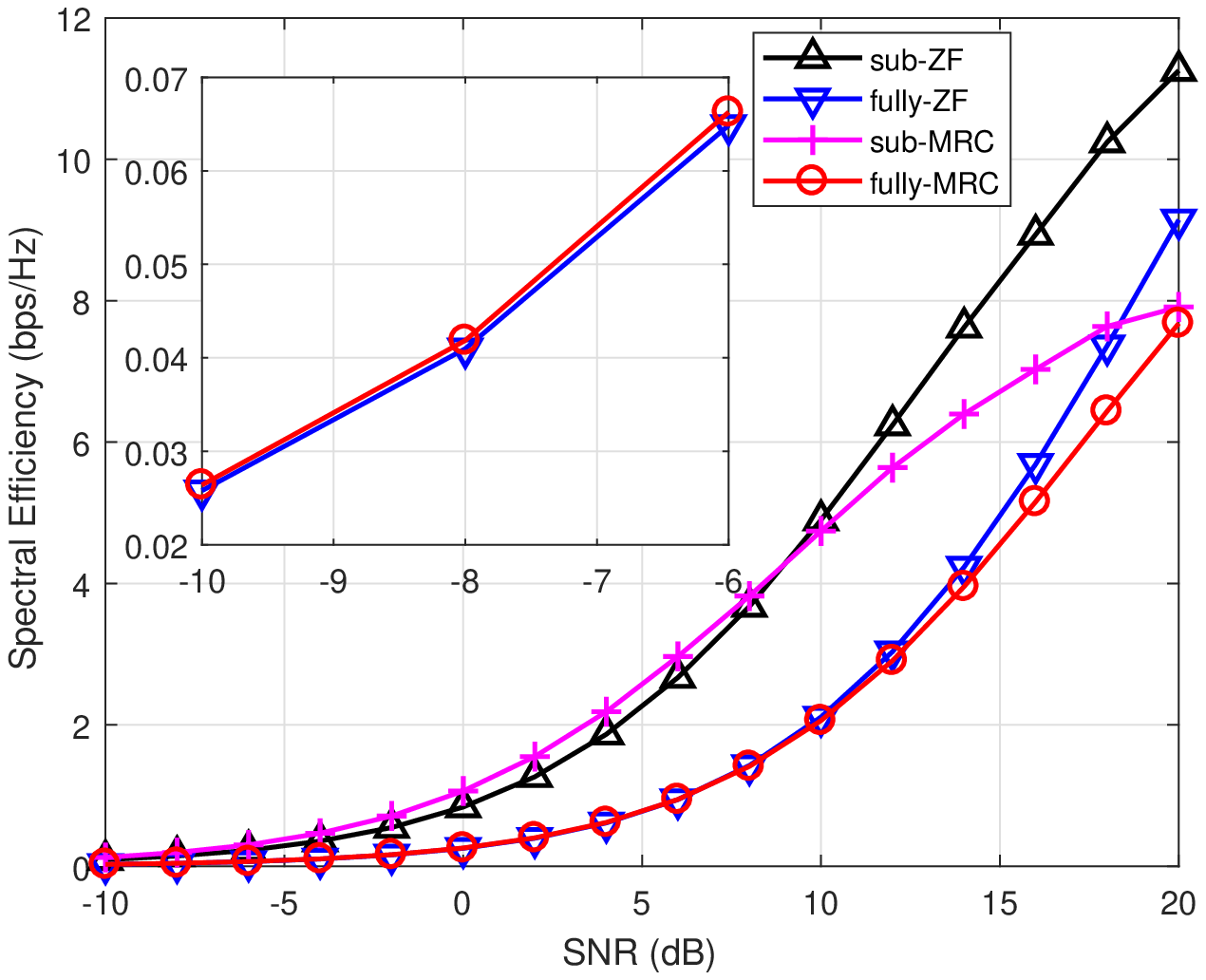}
\caption{Spectral efficiencies using ZF/MRT-based hybrid precoding with $M=64$, $K=4$, $B_1=3$ and $B_2=10$.}
\end{minipage}%
\hfill
\begin{minipage}[c]{0.32\textwidth}
\centering
\includegraphics[width=0.9\textwidth]{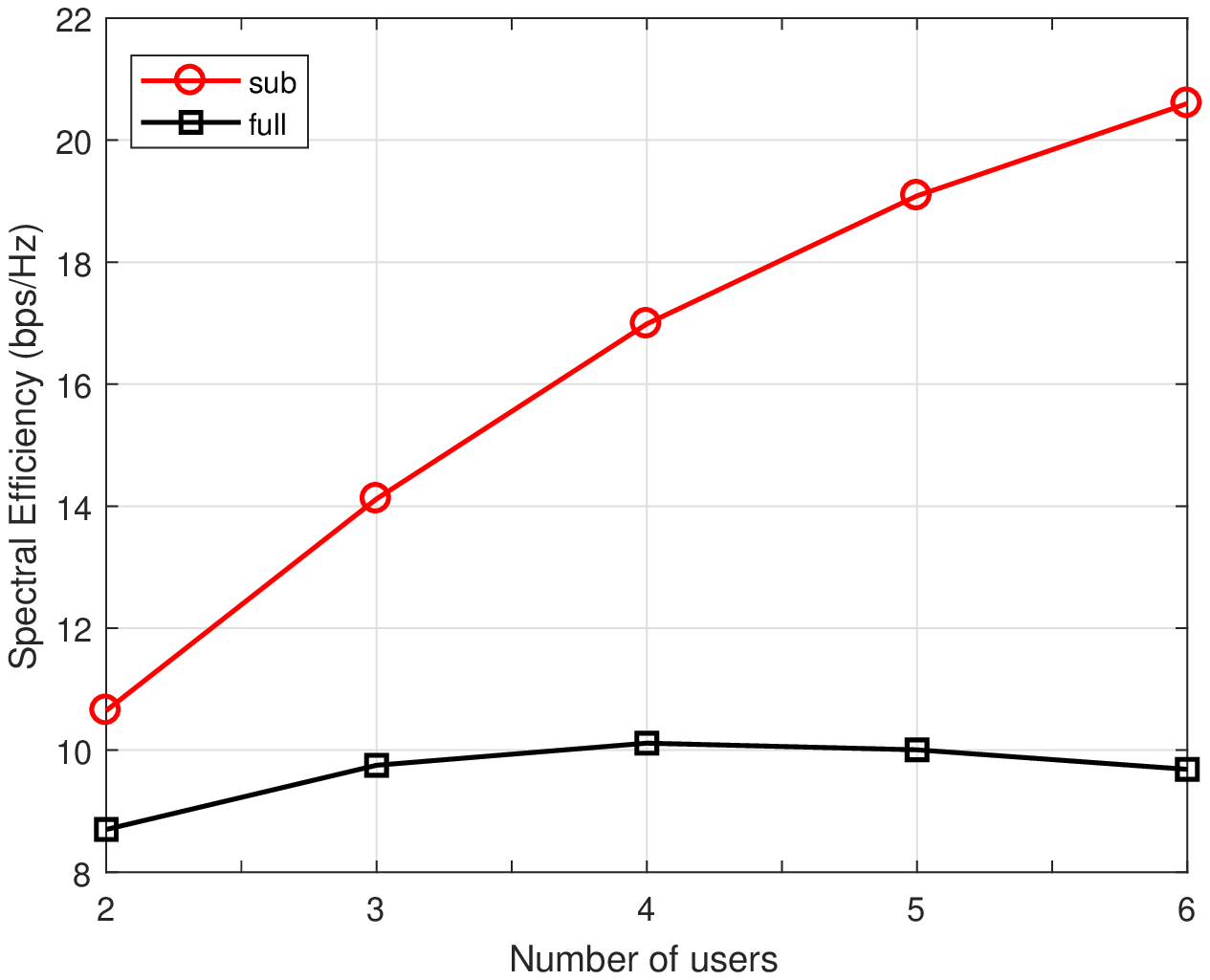}
\caption{Spectral efficiencies using analog quantization with $M=120$, $B_1=3$, SNR$=20$dB.}
\end{minipage}
\end{figure*}

As we know from \eqref{eq25}, the sub-connected structure outperforms the fully-connected structure when perfect quantization is considered. In Fig. 2, we provide results concerning the comparison between the sub-connected hybrid precoding and the fully-digital precoding. For comparison, the fully-digital precoding is considered, where each antenna is connected to one dedicated RF chain without phase shifters, 
splitters or combiners. As shown in Fig. 2, the fully-digital precoding provides better performance as expected. The performance gap due to hybrid precessing becomes larger as the SNR grows. While in the not-too-large SNR region which is of specific interest in (mmWave) massive MIMO scenarios, the gap appears relatively marginal.

In Fig. 3, we provide some simulation results in comparison with hybrid precoding employing maximum ratio transmission (MRT). It is revealed that, at low SNRs, the MRT-based hybrid precoding improves the performance slightly in comparison with ZF-based hybrid precoding
. In constrast, when SNR becomes large enough, the ZF-based hybrid precoding outperforms the MRT-based hybrid precoding. Since the ZF precoding is designed to cope with the interference among users, Lemma 1 shows that the multiuser interference has already been reduced by analog precoding which indicates that the effect of ZF becomes less essential in the hybrid precoding design especially for large antenna numbers. Therefore, the effect of noise boosting in ZF-based hybrid precoding does not affect the system performance significantly.

Simulation results regarding the performance for different numbers of users are provided in Fig. 4. It demonstrates that, for the fully-connected structure, the system spectral efficiency first increases and then decreases with the number of users. While for the sub-connected structure, the system spectral efficiency increases with the number of users.

\begin{figure}[!htpb]
\centering
\begin{minipage}{0.48\textwidth}
\centering
\includegraphics[width=0.75\textwidth]{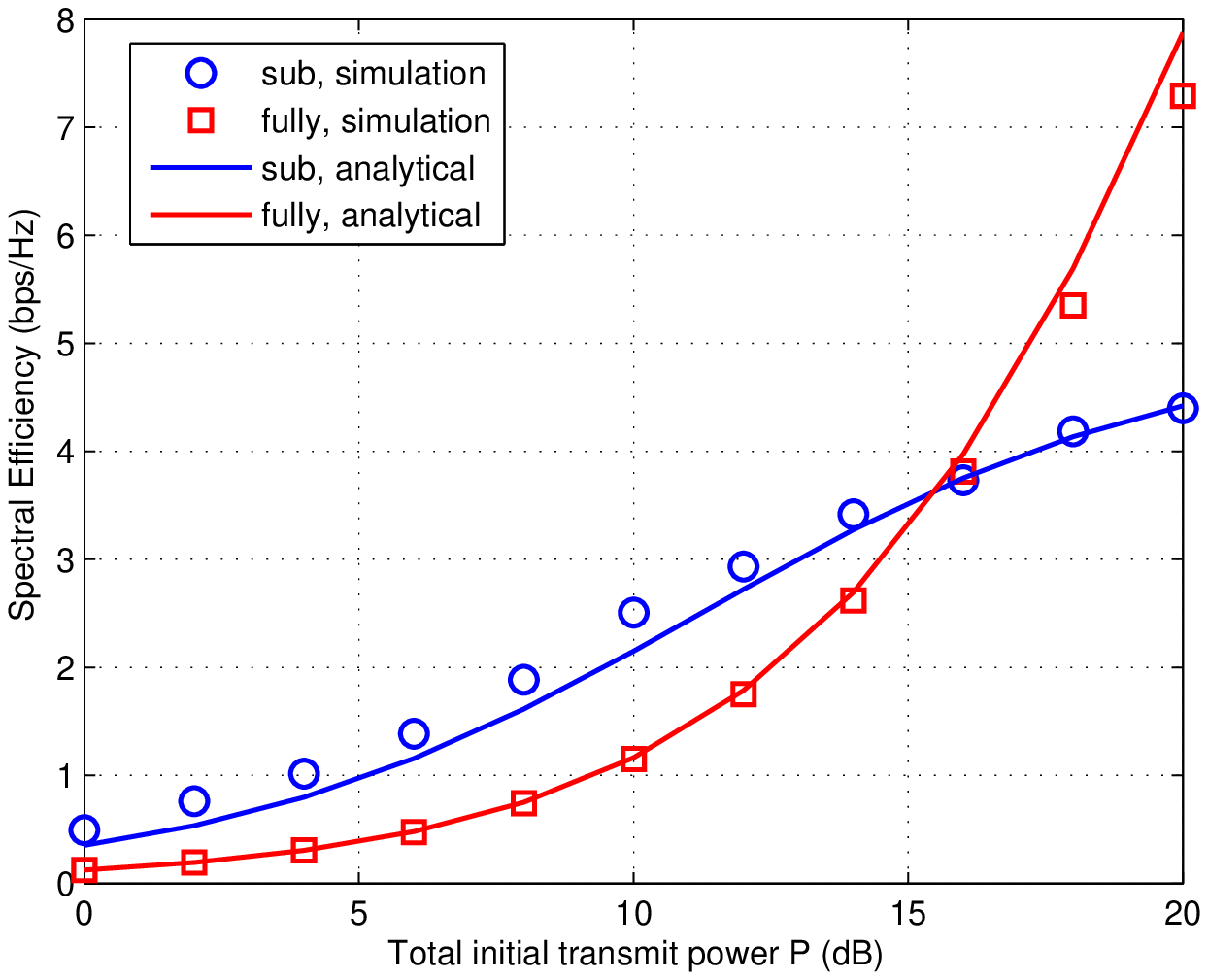}
\caption{Spectral efficiencies using a channel correlation-based codebook with $M=64$, $K=8$, $B_1=3$ and $B_2=6$.}
\end{minipage}%
\hspace{0.1cm}
\begin{minipage}{0.48\linewidth}
\centering
\includegraphics[width=0.75\textwidth]{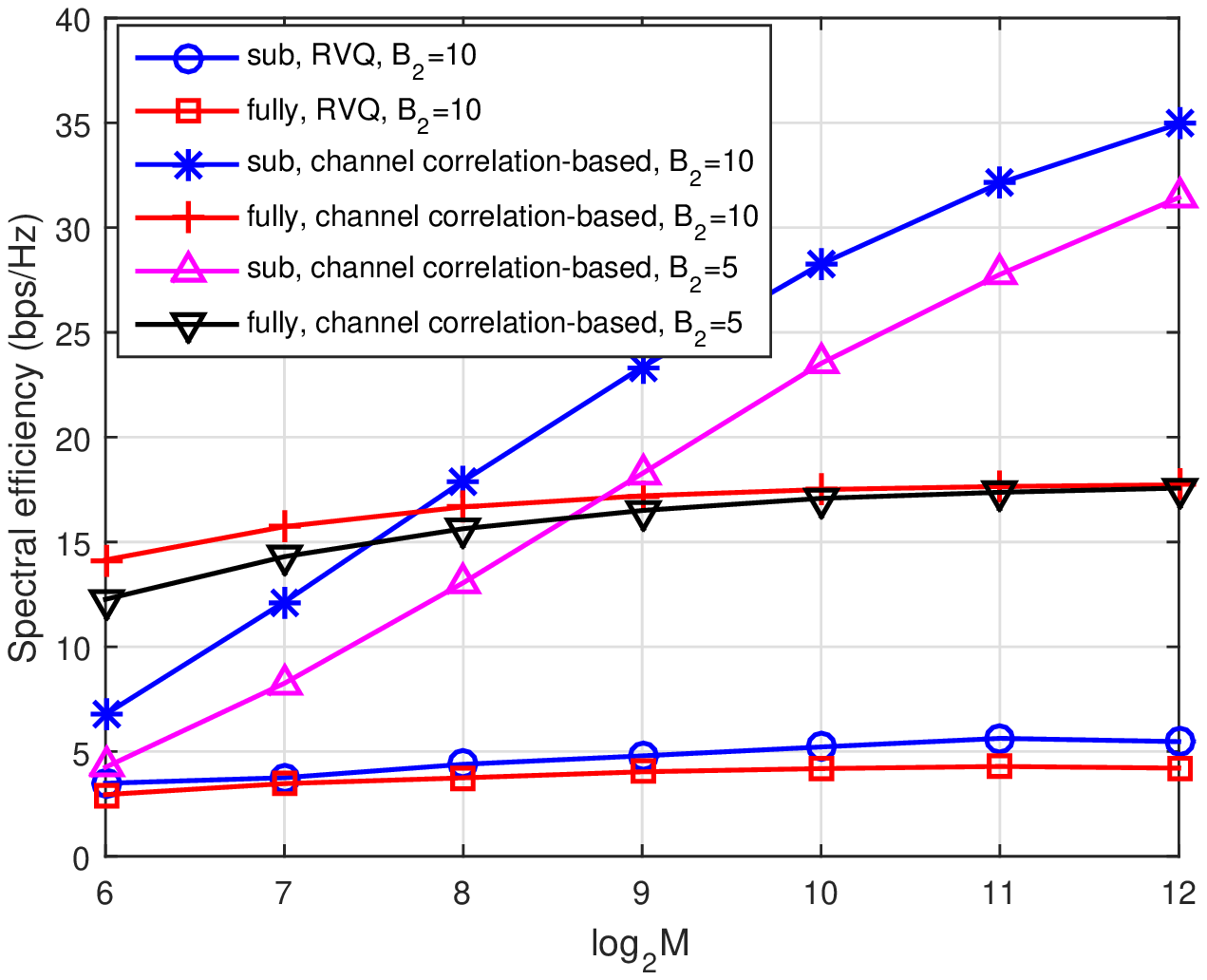}
\caption{Spectral efficiencies using different codebooks with $K=8$, $B_1=3$ and $P=25$dB.}
\end{minipage}
\end{figure}

Fig. 5 displays the spectral efficiencies using the channel correlation-based codebook for different total initial signal powers. It is obvious that the spectral efficiencies derived in Section III are indeed tight. We can also see that, with the same number of antennas, the sub-connected structure has better performance when the SNR is quite low but the spectral efficiency in the fully-connected structure 
surpasses the sub-connected structure as SNR increases.

Fig. 6 displays the spectral efficiencies using different codebooks for the sub-connected and fully-connected structures. Firstly, the sub-connected structure performs worse than the fully-connected structure with small numbers of antennas but surpasses the fully-connected structure with a large $M$. 
Then, the channel correlation-based codebook is preferred since it achieves a far larger spectral efficiency than RVQ. Especially, the system using the channel correlation-based codebook with $B_2=5$ outperforms the one using RVQ with $B_2=10$. Thereby, the channel correlation-based codebook enables hybrid precoding systems to achieve better performance with fewer feedback bits. Besides, only in the sub-connected structure using the channel correlation-based codebook, the performance improves remarkably with the increase in $M$.

\subsection{mmWave Multiuser Channels}

Rayleigh fading is popularly used, but somewhat simplified for characterizing the random nature of wireless channels which is more tractable for analysis and getting insightful observations. Apart from 
Rayleigh 
channels, hybrid precoding using the channel correlation-based codebook can also be applied to mmWave communications. For mmWave channels
, the geometric channel model is currently leveraged as a more accurate model as discussed in \cite{6717211} \cite{6387266}. Since this model involves practical scattering features, it becomes less tractable in conducting theoretical analysis with engineering insights. Therefore, we consider Rayleigh 
channels during the derivations and provide some simulation results regarding the mmWave channels in this subsection.

We assume that each user has the same number of scatters and each scatter contributes to a single propagation path between the BS and the user. For the single-antenna user, the channel model can be expressed as
$\mathbf{h}_k^H=\sqrt{\frac{M}{L}}\sum_{l=1}^L\alpha_l^k\mathbf{a}_{BS}^H(\phi_l^k)$
where $L$ denotes the number of propagation paths from BS to each user, $\alpha_l^k\sim\mathcal{CN}(0,1)$ represents the complex gain of the $l$-th path. The variable $\phi_l^k$ is the $l$-th path's azimuth angle of departure which follows uniform distribution over [$0$,$2\pi$). Finally, $\mathbf{a}_{BS}^H(\phi_l^k)$ is the antenna array response vector of the BS which is only dependent on specific array structures. For simplicity, we utilize uniform linear arrays (ULAs) in the simulation, under which $\mathbf{a}_{BS}^H(\phi_l^k)$ can be defined as $\mathbf{a}_{BS}^H(\phi_l^k)=\frac{1}{\sqrt{M}}\left[1,e^{j\frac{2\pi}{\lambda}d\sin(\phi_l^k)},...,e^{j(M-1)\frac{2\pi}{\lambda}d\sin(\phi_l^k)}\right]^T$ where $\lambda$ is the signal wavelength, and $d$ is the distance between antenna elements.

\begin{figure}[!htpb]
\centering
\begin{minipage}{0.48\textwidth}
\centering
\includegraphics[width=0.75\textwidth]{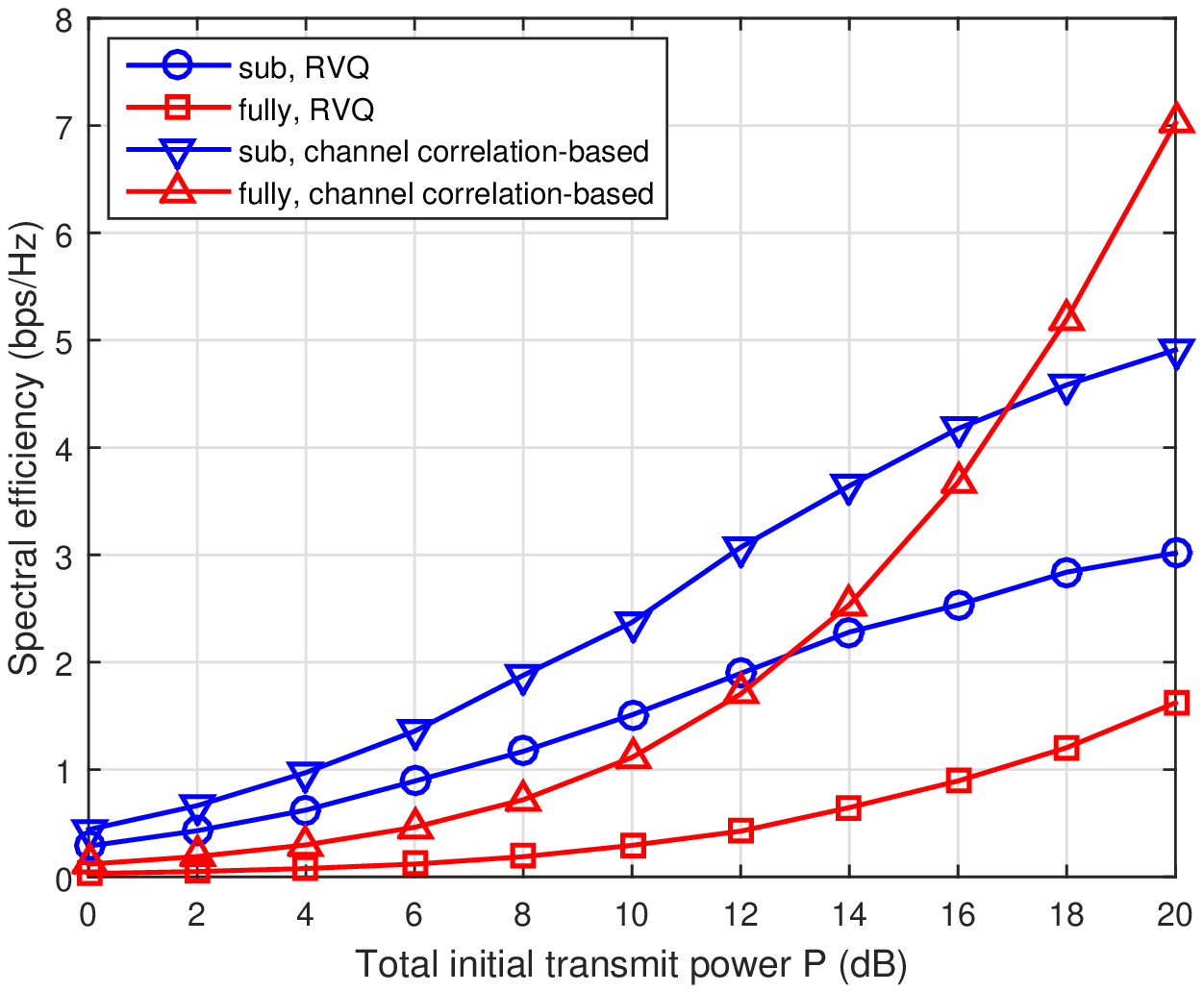}
\caption{Spectral efficiencies using different codebooks over mmWave channels with $L=10$ and $\frac{d}{\lambda}=\frac{1}{2}$.}
\end{minipage}
\hspace{0.1cm}
\begin{minipage}{0.48\textwidth}
\centering
\includegraphics[width=0.75\textwidth]{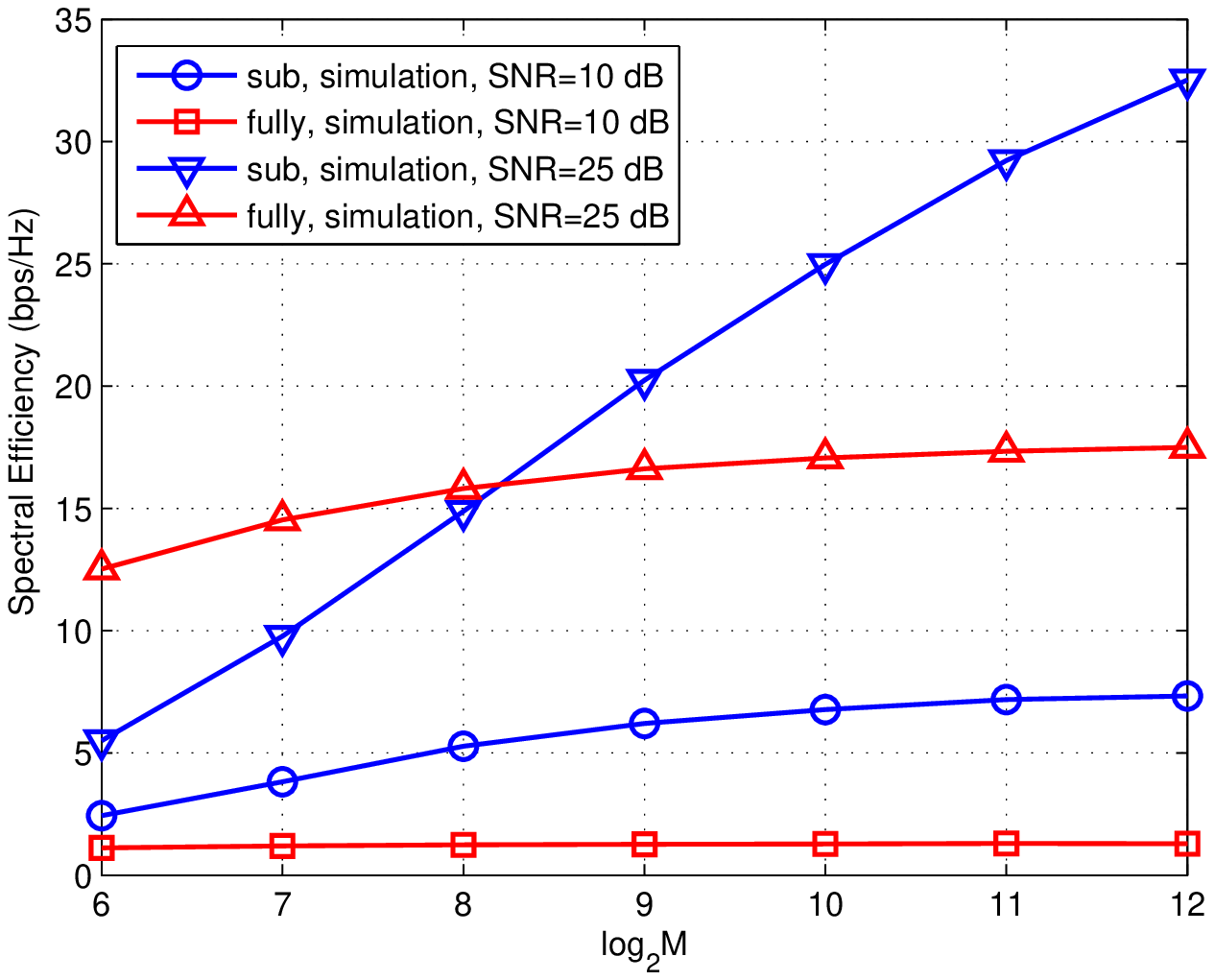}
\caption{Spectral efficiencies using a channel correlation-based codebook over mmWave channels with $L=10$ and $\frac{d}{\lambda}=\frac{1}{2}$.}
\end{minipage}
\end{figure}

We compare the two quantized digital codebooks over poor scattering mmWave channels in Fig. 7. We consider the same setting as Fig. 5 despite $B_2=10$. Apparently, the channel correlation-based codebook achieves a higher spectral efficiency than the RVQ codebook. In addition, the performance of different structures and different codebooks increases with the total initial signal power
while the performance gap between the two codebooks also grows with the total initial signal power in both structures.

Fig. 8 displays the effect of total initial signal power and number of BS antennas on spectral efficiencies using the channel correlation-based codebook 
over mmWave channels. We adopt the same setting as Fig. 6. It is shown that the sub-connected structure always has better performance at low SNRs, while for relatively high SNRs, it exhibits worse performance with a small $M$ but outperforms the fully-connected structure when $M$ is large enough.

\section{Conclusion}

In this paper, we have studied the massive multiuser MIMO systems over limited feedback channels under a realistic hardware network model with dissipation. A channel correlation-based codebook has been employed according to the effective channels in hybrid processing. We have also compared the spectral efficiencies of hybrid precoding for the sub-connected structure and the fully-connected structure. Analytical and simulation results show that the system spectral efficiency is better in the sub-connected structure than in the more complex fully-connected structure in systems with a massive antenna array or low SNR; otherwise, the fully-connected structure achieves better performance. Furthermore, the channel correlation-based codebook outperforms the conventional RVQ codebook in hybrid precoding systems.

\begin{appendices}

\section{Proof of Lemma 1}

We first prove the part regarding the sub-connected structure. From \eqref{eq22}, we have
\begin{align}
g_{k,k}&=(\mathbf{h}_k^H\hat{\mathbf{a}}_k)^*
\overset{(a)}{=}\frac{1}{N}\sum_{i=(k-1)N+1}^{kN}h_{k,i}{\rm e}^{-j\hat{\varphi}_{k,i}}
=\frac{1}{N}\sum_{i=(k-1)N+1}^{kN}\lambda_i\label{eq47}
\end{align}
where (a) uses \eqref{eq02}, and $\lambda_i\overset{\triangle}{=}h_{i,k}{\rm e}^{-j\hat{\varphi}_{k,i}}$. Define $\varepsilon_{k,i}$ as the phase error between the quantized phase $\varphi_{k,i}$ and quantized phase $\hat{\varphi}_{k,i}$, i.e., $\varepsilon_{k,i}\overset{\triangle}{=}\varphi_{k,i}-\hat{\varphi}_{k,i}$. It yields $\lambda_i=|h_{i,k}|{\rm e}^{j\varepsilon_{k,i}}$. Since ${\rm e}^{j\varepsilon_{k,i}}$ and $|h_{i,k}|$ are independent random variables, we investigate them separately in the following.

As the phase of each entry in channel matrix $\mathbf{H}$ follows uniform distribution between $0$ and $2\pi$, i.e., $\varphi_{k,j}\sim U[0,2\pi)$, we can easily conclude the distribution of the phase error as $\varepsilon_{k,i}\sim U [-\delta,\delta)$ where we define $\delta\overset{\triangle}{=}\frac{\pi}{2^{B_1}}$. Then, according to the Euler's formula, it is obtained that
\begin{align}
\mathbb{E}[\Re[{\rm e}^{j\varepsilon_{k,i}}]]&=\frac{1}{2\delta}\int_{-\delta}^\delta\cos\varepsilon_{k,i}{\rm d}\varepsilon_{k,i}={\rm sinc}(\delta)\label{eq40}
\end{align}
and
\begin{align}
\mathbb{E}[(\Re[{\rm e}^{j\varepsilon_{k,i}}])^2]&=\frac{1}{2\delta}\int_{-\delta}^\delta\cos^2\varepsilon_{k,j}{\rm d}\varepsilon_{k,j}=\frac{1}{2}[1+{\rm sinc}(\delta)\cos(\delta)]\label{eq41}
\end{align}
where ${\rm sinc}(\delta)=\frac{\sin(\delta)}{\delta}$. Recalling $\mathbf{h}_k\sim \mathcal{CN}(\mathbf{0}_M,\mathbf{I}_M)$, $|h_{i,k}|$ follows the Rayleigh distribution and hence
\begin{align}
\mathbb{E}[|h_{i,k}|]=&\frac{\sqrt{\pi}}{2},\
\mathbb{V}[|h_{i,k}|]=1-\frac{\pi}{4},\
\mathbb{E}[|h_{i,k}|^2]=\mathbb{E}^2[|h_{i,k}|]+\mathbb{V}[|h_{i,k}|]=1.\label{eq43}
\end{align}

Since $\lambda_i=|h_{i,k}|{\rm e}^{j\varepsilon_{k,i}}$, it is obvious that
$\Re[\lambda_i]=|h_{i,k}|\Re[{\rm e}^{j\varepsilon_{k,i}}]$ and
$\Im[\lambda_i]=|h_{i,k}|\Im[{\rm e}^{j\varepsilon_{k,i}}]$.
Owing to the dependence between $|h_{i,k}|$ and ${\rm e}^{j\varepsilon_{k,i}}$, we can further have
\begin{align}
\mathbb{E}[\Re[\lambda_i]]&=\mathbb{E}[|h_{i,k}|]\mathbb{E}[\Re[{\rm e}^{j\varepsilon_{k,i}}]]=\frac{\sqrt{\pi}}{2}{\rm sinc}(\delta)\label{eq44}\\
\mathbb{E}[(\Re[\lambda_i])^2]&=\mathbb{E}[(|[h_{i,k}|\Re[{\rm e}^{j\varepsilon_{k,i}}])^2]
=\mathbb{E}[|h_{i,k}^*|^2]\mathbb{E}[(\Re[{\rm e}^{j\varepsilon_{k,i}}])^2]
=\frac{1}{2}[1+{\rm sinc}(\delta)\cos(\delta)]\label{eq45}
\end{align}
where \eqref{eq40}-\eqref{eq43} are used in the last equality.
According to \eqref{eq44} and \eqref{eq45}, it gives
\begin{align}
\mathbb{V}[\Re[\lambda_i]]=&\mathbb{E}[(\Re[\lambda_i])^2]-(\mathbb{E}[\Re[\lambda_i]])^2=\omega_1\label{eq46}
\end{align}
where $\omega_1=\frac{1}{2}[1+{\rm sinc}(\delta)\cos(\delta)]-\frac{\pi}{4}{\rm sinc}^2(\delta)$.
By applying the Central Limit Theorem to \eqref{eq47} and utilizing \eqref{eq44} and \eqref{eq46}, we get
\begin{align}
\Re[g_{k,k}]\sim\mathcal{N}\left(\frac{\sqrt{\pi}}{2}{\rm sinc}(\delta),\frac{\omega_1}{N}\right)\label{eq48}
\end{align}
for the massive MIMO with large $N$. Similarly, we can show that, for large $N$,
\begin{align}
\Im[g_{k,k}]\sim\mathcal{N}\left(0,\frac{\omega_2}{N}\right)\label{eq49}
\end{align}
where $\omega_2=\frac{1}{2}[1-{\rm sinc}(\delta)\cos(\delta)]$. It is obvious that
\begin{align}
\lim_{N\rightarrow\infty}\frac{\omega_1}{N}=&0,\
\lim_{N\rightarrow\infty}\frac{\omega_2}{N}=0.\label{eq50}
\end{align}
As $N\rightarrow\infty$ in massive MIMO, according to \eqref{eq48}, \eqref{eq49} and \eqref{eq50}, we obtain
\begin{align}
\Re[g_{k,k}]\overset{a.s.}{\rightarrow}&\frac{\sqrt{\pi}}{2}{\rm sinc}(\delta),\
\Im[g_{k,k}]\overset{a.s.}{\rightarrow}0.\label{eq51}
\end{align}

Following trivially the steps above, it yields
\begin{align}
\mathbb{E}[\Re[h_{i,k}{\rm e}^{j\hat{\varphi}_{k,i}}]]=&\mathbb{E}[\Im[h_{i,k}{\rm e}^{j\hat{\varphi}_{k,i}}]]=0,\
\mathbb{V}[\Re[h_{i,k}{\rm e}^{j\hat{\varphi}_{k,i}}]]=\mathbb{V}[\Im[h_{i,k}{\rm e}^{j\hat{\varphi}_{k,i}}]]=\frac{1}{2}.\label{eq66}
\end{align}
Using the the Central Limit Theorem to $g_{k,j}=\frac{1}{N}\sum_{i=(k-1)N+1}^{kN}h_{k,i}{\rm e}^{-j\hat{\varphi}_{j,i}}$, we have
\begin{align}
\Re[g_{k,j}]\sim&\mathcal{N}\left(0,\frac{1}{2N}\right),\
\Im[g_{k,j}]\sim\mathcal{N}\left(0,\frac{1}{2N}\right)\label{eq55}
\end{align}
which implies for large $N$ that
\begin{align}
\Re[g_{k,j}]\overset{a.s.}{\rightarrow}&0,\
\Im[g_{k,j}]\overset{a.s.}{\rightarrow}0.\label{eq52}
\end{align}
Combining \eqref{eq51} and \eqref{eq52}, we consequently prove the part concerning the sub-connected structure.

The proof for the fully-connected structure is 
 analogous to that for the sub-connected structure. We omit it for brevity.

\section{Proof of Lemma 2}

For the sub-connected structure, according to \eqref{eq27}, investigating the correlation matrix amounts to calculating every element, i.e., $r_{i,j}$, in the matrix. The calculation of $r_{i,j}$ can be separately conducted for three cases: (1) $i=j=k$; (2) $i=j\neq k$; (3) $i\neq j$. We will evaluate case-by-case for the sub-connected structure:

(1) $i=j=k$: From the definition in \eqref{eq27}, we obtain $r_{k,k}=\mathbb{E}[g_{k,k}g_{k,k}^*]
=\mathbb{E}[\Re^2[g_{k,k}]+\Im^2[g_{k,k}]]
=\frac{\pi}{4}{\rm sinc}^2(\delta)+\frac{\omega_1}{N}+\frac{\omega_2}{N}$
where we recall that $\Re[g_{k,k}]\sim\mathcal{N}\left(\rm {sinc}(\delta)\frac{\sqrt{\pi }}{2},\frac{\omega_1}{N}\right)$ and $\Im[g_{k,k}]\sim\mathcal{N}\left(0,\frac{\omega_2}{N}\right)$ in Appendix A.

(2) $i=j\neq k$: Similarly, using $r_{i,j}=\mathbb{E}[\Re^2[g_{k,i}]+\Im^2[g_{k,i}]]$ and \eqref{eq55}, we get $r_{i,i}=\frac{1}{N}, (i\neq k)$.

(3) $i\neq j$: If $i\neq k$ and $j\neq k$, from the definition in \eqref{eq27}, we have
\begin{align}
r_{i,j}=&\mathbb{E}[g_{k,i}g_{k,j}^*]\nonumber\\
=&\mathbb{C}ov[g_{k,i},g_{k,j}]+\mathbb{E}[g_{k,i}]\mathbb{E}[g_{k,j}^*]\nonumber\\
\overset{(a)}{=}&\frac{1}{2}(\mathbb{V}[g_{k,i}+g_{k,j}]-\mathbb{V}[g_{k,i}]-\mathbb{V}[g_{k,j}])+\mathbb{E}[g_{k,i}]\mathbb{E}[g_{k,j}^*]\nonumber\\
=&\frac{1}{2}(\mathbb{V}[\kappa_k]-\mathbb{V}[g_{k,i}]-\mathbb{V}[g_{k,j}])+\mathbb{E}[g_{k,i}]\mathbb{E}[g_{k,j}^*]\label{eq54}
\end{align}
where (a) follows from $\mathbb{C}ov[g_{k,i},g_{k,j}]=\mathbb{C}ov[g_{k,j},g_{k,i}]$ and the definition of covariance, and
\begin{align}
\kappa_k&\overset{\triangle}{=}g_{k,i}+g_{k,j}
=\mathbf{h}_k^H\hat{\mathbf{a}}_i+\mathbf{h}_k^H\hat{\mathbf{a}}_j
=\frac{1}{N}\left(\sum_{l=N(i-1)+1}^{Ni}h_{l,k}^*e^{\hat{\varphi}_{i,l}}+\sum_{l=N(j-1)+1}^{Nj}h_{l,k}^*e^{\hat{\varphi}_{j,l}}\right)
=\frac{1}{N}\sum_{l=1}^{2N}\chi_l\label{eq56}
\end{align}
in which $\chi_l$ is defined as
\begin{align}
\chi_l\overset{\triangle}{=}\left\{\begin{array}{ll}
h_{l+N(i-1),k}^*e^{\hat{\varphi}_{i,l+N(i-1)}}, & 1\leq l\leq N\\
h_{l+N(j-2),k}^*e^{\hat{\varphi}_{j,l+N(j-2)}}, & N<l\leq2N.
\end{array}\right.\label{eq39}
\end{align}
By using \eqref{eq55}, it is not difficult to get
\begin{align}
\mathbb{V}[g_{k,i}]=&\mathbb{V}[g_{k,j}]
=\mathbb{E}[|g_{k,i}|^2]-|\mathbb{E}[g_{k,i}]|^2\nonumber\\
=&\mathbb{E}[(\Re[g_{k,i}])^2]+\mathbb{E}[(\Im[g_{k,i}])^2]-|\mathbb{E}[\Re[g_{k,i}]+\sqrt{-1}\Im[g_{k,i}]]|^2\nonumber\\
=&\mathbb{V}[\Re[g_{k,i}]]+(\mathbb{E}[\Re[g_{k,i}])^2+\mathbb{V}[\Im[g_{k,i}]]+(\mathbb{E}[\Im[g_{k,i}])^2-|\mathbb{E}[\Re[g_{k,i}]+\sqrt{-1}\Im[g_{k,i}]]|^2\nonumber\\
=&\frac{1}{N}\label{eq57}
\end{align}
As we have already obtained $\mathbb{V}[g_{k,i}]$ and $\mathbb{V}[g_{k,j}]$, the remaining work is evaluating $\kappa_k$. In order to analyze $\kappa_k$, according to \eqref{eq56}, we need to investigate $\chi_l$. Combining \eqref{eq66} and \eqref{eq39}, it gives
\begin{align}
\mathbb{E}[\Re[\chi_l]]=&\mathbb{E}[\Im[\chi_l]]=0,\
\mathbb{V}[\Re[\chi_l]]=\mathbb{V}[\Im[\chi_l]]=\frac{1}{2}.\label{eq67}
\end{align}
Applying the Central Limit Theorem and using \eqref{eq56} and \eqref{eq67}, it yields
\begin{align}
\Re[\kappa_k]&\sim\mathcal{N}\left(0,\frac{1}{N}\right),\
\Im[\kappa_k]\sim\mathcal{N}\left(0,\frac{1}{N}\right)\label{eq37}
\end{align}
which indicates that
\begin{align}
\mathbb{E}[(\Re[\kappa_k])^2]=\mathbb{V}[\Re[\kappa_k]]+\mathbb{E}^2[\Re[\kappa_k]]&=\frac{1}{N},\
\mathbb{E}[(\Im[\kappa_k])^2]=\mathbb{V}[\Im[\kappa_k]]+\mathbb{E}^2[\Im[\kappa_k]]&=\frac{1}{N}.\label{eq38}
\end{align}
Therefore,we have
\begin{align}
\mathbb{V}[\kappa_k]=\mathbb{E}[|\kappa_k|^2]-|\mathbb{E}[\kappa_k]|^2
=\mathbb{E}[(\Re[\kappa_k])^2]+\mathbb{E}[(\Im[\kappa_k])^2]-|\mathbb{E}[\Re[\kappa_k]+\sqrt{-1}\Im[\kappa_k]]|^2
=\frac{2}{N}\label{eq53}
\end{align}
where \eqref{eq37} and \eqref{eq38} are utilized. Substituting \eqref{eq57} and \eqref{eq53} into \eqref{eq54} and using \eqref{eq55}, we obtain $r_{i,j}=0$.

If $i=k$ or $j=k$, the proof is similar with the case above when $i\neq k$ and $j\neq k$. Due to the brevity, we leave it out here.

Summarizing the three cases above, we prove the part concerning the sub-connected structure.

The part for the fully-connected structure can be proved in an analogous fashion.

\section{Proof of Theorem 2}

We first discuss the part for the sub-connected structure. The part regarding the fully-connected one is similar and we leave it out for brevity.

From \eqref{eq29}, the spectral efficiency loss for the sub-connected structure can be expressed as
\begin{align}
\Delta R_{sub}=&R_k^{sub}-\bar{R}_{sub}^Q\nonumber\\
=&\mathbb{E}\left[\log_2\left(1+\frac{P}{K}|\mathbf{g}_k^H\mathbf{w}_k|^2\right)\right]
-\mathbb{E}\left[\log_2\left(1+\frac{\frac{P}{K}|\mathbf{g}_k^H\hat{\mathbf{w}}_k|^2}{\frac{P}{K}\sum_{j\neq k}^K|\mathbf{g}_k^H\hat{\mathbf{w}}_j|^2+1}\right)\right]\nonumber\\
=&\mathbb{E}\left[\log_2\left(1+\frac{P}{K}|\mathbf{g}_k^H\mathbf{w}_k|^2\right)\right]
-\mathbb{E}\left[\log_2\left(1+\frac{P}{K}|\mathbf{g}_k^H\hat{\mathbf{w}}_k|^2+\frac{P}{K}\sum_{j\neq k}^K|\mathbf{g}_k^H\hat{\mathbf{w}}_j|^2\right)\right]\nonumber\\
&+\mathbb{E}\left[\log_2\left(1+\frac{P}{K}\sum_{j\neq k}^K|\mathbf{g}_k^H\hat{\mathbf{w}}_j|^2\right)\right]\nonumber\\
\overset{(a)}{\leq}&\mathbb{E}\left[\frac{\log_2\left(1+\frac{P}{K}|\mathbf{g}_k^H\mathbf{w}_k|^2\right)}{\log_2\left(1+\frac{P}{K}|\mathbf{g}_k^H\hat{\mathbf{w}}_k|^2\right)}\right]
+\mathbb{E}\left[\log_2\left(1+\frac{P}{K}\sum_{j\neq k}^K|\mathbf{g}_k^H\hat{\mathbf{w}}_j|^2\right)\right]\nonumber\\
\overset{(b)}{\leq}&\mathbb{E}\left[\log_2\left(\frac{\|\mathbf{g}_k^H\|^2}{\|\mathbf{g}_k^H\|^2}\right)\right]+\mathbb{E}\left[\log_2\left(\frac{P}{K}|\tilde{\mathbf{g}}_k^H\mathbf{w}_k|^2\right)\right]
-\mathbb{E}\left[\log_2\left(\frac{P}{K}|\tilde{\mathbf{g}}_k^H\hat{\mathbf{w}}_k|^2\right)\right]\nonumber\\
&+\mathbb{E}\left[\log_2\left(1+\frac{P}{K}\sum_{j\neq k}^K|\mathbf{g}_k^H\hat{\mathbf{w}}_j|^2\right)\right]\label{eq58}
\end{align}
where (a) results from removing the positive quantity $\frac{P}{K}\sum_{j\neq k}^K|\mathbf{g}_k^H\hat{\mathbf{w}}_j|^2$ from the second term, and (b) is due to the fact that $\log(\frac{1+a}{1+b})\leq\log(\frac{a}{b})$ holds for any positive numbers $a$ and $b$, and $\tilde{\mathbf{g}}_k=\frac{\mathbf{g}_k}{\|\mathbf{g}_k\|}$ is the normalized effective channel. In addition, as the ZF precoding vectors $\mathbf{w}_k$ and $\hat{\mathbf{w}}_k$ are designed to be in the null space of the other users' channel vectors, they are independent of $\tilde{\mathbf{g}}_k$. Therefore, $\mathbb{E}\left[\log_2\left(\frac{P}{K}|\tilde{\mathbf{g}}_k^H\mathbf{w}_k|^2\right)\right]$ and $\mathbb{E}\left[\log_2\left(\frac{P}{K}|\tilde{\mathbf{g}}_k^H\hat{\mathbf{w}}_k|^2\right)\right]$ are equal which indicates
\begin{align}
\Delta R_{sub}=&\mathbb{E}\left[1+\log_2\left(\frac{P}{K}\sum_{j\neq k}^K|\mathbf{g}_k^H\hat{\mathbf{w}}_j|^2\right)\right]\nonumber\\
\overset{(a)}{\leq}&\log_2\left(1+\frac{P(K-1)}{K}\mathbb{E}\left[\|\mathbf{g}_k\|^2\right]\mathbb{E}\left[|\tilde{\mathbf{g}}_k^H\hat{\mathbf{w}}_j|^2\right]\right)\label{eq59}\\
\overset{(b)}{\leq}&\log_2\left(1+\frac{P(K-1)}{K}\mathbb{E}\left[\|\mathbf{g}_k\|^2\right]\mathbb{E}\left[1-|\tilde{\mathbf{g}}_k^H\hat{\mathbf{g}}_k|^2\right]\right)\label{eq11}
\end{align}
where (a) follows from Jensen's inequality and (b) is due to the orthogonality between $\hat{\mathbf{g}}_k$ and $\hat{\mathbf{w}}_j$ that $\|\mathbf{g}_k\|^2\geq|\tilde{\mathbf{g}}_k^H\hat{\mathbf{w}}_j|^2+\|\mathbf{g}_k\|^2|\tilde{\mathbf{g}}_k^H\hat{\mathbf{g}}_k|^2$.
The remaining work is to investigate $\mathbb{E}\left[\|\mathbf{g}_k\|^2\right]$ and $\mathbb{E}\left[1-|\tilde{\mathbf{g}}_k^H\hat{\mathbf{g}}_k|^2\right]$ respectively which is given in the following.

From Appendix A, we have $\Re[g_{k,k}]\sim\mathcal{N}\left(\frac{\sqrt{\pi}}{2}{\rm sinc}(\delta),\frac{\omega_1}{N}\right)$ and $\Im[g_{k,k}]\sim\mathcal{N}\left(0,\frac{\omega_2}{N}\right)$, while for the off-diagonal terms, $\Re[g_{k,i}]\sim\mathcal{N}\left(0,\frac{1}{2N}\right)$ and $\Im[g_{k,i}]\sim\mathcal{N}\left(0,\frac{1}{2N}\right)$. Hence,
\begin{align}
\mathbb{E}\left[\|\mathbf{g}_k\|^2\right]&=\mathbb{E}\left[|g_{k,k}|^2\right]+\mathbb{E}\left[\sum_{j=1,j\neq k}^K|g_{k,j}|^2\right]
=\frac{\pi}{4}{\rm sinc}^2(\delta)+\frac{\omega_1}{N}+\frac{\omega_2}{N}+\frac{K-1}{N}. \label{eq04}
\end{align}
Then, we turn to derive
$\mathbb{E}\left[1-|\tilde{\mathbf{g}}_k^H\hat{\mathbf{g}}_k|^2\right]$. According to \cite{4697964}, we can write $\mathbb{E}\left[1-|\tilde{\mathbf{g}}_k^H\hat{\mathbf{g}}_k|^2\right]\approx \frac{\sigma_{k,2}^2}{\sigma_{k,1}^2}2^{-\frac{B_2}{K-1}}$
where $\sigma_{k,1}$ 
is the largest singular value of $\mathbf{R}_{k}^{1/2}$ and $\sigma_{k,2}$ is the second largest singular value of $\mathbf{R}_{k}^{1/2}$.

For the sub-connected structure, $\mathbf{R}_{k}$ is diagonal. Combining \eqref{eq12} and \eqref{eq09}, we can further obtain $\sigma_{k,1}^2=\frac{\pi}{4}{\rm sinc}^2(\delta)+\frac{\omega_1}{N}+\frac{\omega_2}{N}$ and $\sigma_{k,2}^2=\frac{1}{N}$. Therefore,
\begin{align}
\mathbb{E}\left[1-|\tilde{\mathbf{g}}_k^H\hat{\mathbf{g}}_k|^2\right]\approx \frac{1}{\frac{\pi N}{4}{\rm sinc}^2(\delta)+\omega_1+\omega_2}2^{-\frac{B_2}{K-1}}\label{eq10}.
\end{align}

Substituting \eqref{eq04} and \eqref{eq10} into \eqref{eq11}, we have
\begin{align}
\Delta R_{sub}\lesssim&\log_2\left(1+\frac{P(K-1)}{K}\frac{1}{\frac{\pi N}{4}{\rm sinc}^2(\delta)+\omega_1+\omega_2}
\left[\frac{\pi}{4}{\rm sinc}^2(\delta)+\frac{\omega_1}{N}+\frac{\omega_2}{N}+\frac{K-1}{N}\right]2^{-\frac{B_2}{K-1}}\right)\nonumber\\
\overset{(a)}\rightarrow&\log_2\left(1+\frac{P(K-1)}{M}2^{-\frac{B_2}{K-1}}\right)
\end{align}
where (a) holds because $\frac{\frac{\pi N}{4}{\rm sinc}^2(\delta)+\omega_1+\omega_2+K-1}{\frac{\pi N}{4}{\rm sinc}^2(\delta)+\omega_1+\omega_2}\rightarrow1$ as $N\rightarrow\infty$. It completes the proof concerning the sub-connected structure.

\section{Proof of \eqref{eq31}}

We first discuss the part for the sub-connected structure. The part concerning the fully-connected structure can be proved readily in a similar manner and we omit it for brevity. Trivially following steps in \eqref{eq58} and \eqref{eq59}, it yields
\begin{align}
\Delta R_{sub}^{RVQ}\leq&\log_2\left(1+\frac{P(K-1)}{K}\mathbb{E}\left[\|\mathbf{g}_k\|^2\right]\mathbb{E}\left[|\tilde{\mathbf{g}}_k^H\hat{\mathbf{w}}_j^{RVQ}|^2\right]\right)\nonumber\\
\overset{(a)}{=}&\log_2\left(1+\frac{P(K-1)}{K}\mathbb{E}\left[|\tilde{\mathbf{g}}_k^H\hat{\mathbf{w}}_j^{RVQ}|^2\right]
\left[\frac{\pi}{4}{\rm sinc}^2(\delta)+\frac{\omega_1}{N}+\frac{\omega_2}{N}+\frac{K-1}{N}\right]\right)\nonumber\\
\overset{(b)}{\rightarrow}&\log_2\left(1+\frac{\pi P(K-1)}{4K}{\rm sinc}^2(\delta)\mathbb{E}\left[|\tilde{\mathbf{g}}_k^H\hat{\mathbf{w}}_j^{RVQ}|^2\right]\right)\label{eq63}
\end{align}
where (a) uses \eqref{eq04}, (b) follows from $\frac{\frac{\pi}{4}{\rm sinc}^2(\delta)+\frac{\omega_1}{N}+\frac{\omega_2}{N}+\frac{K-1}{N}}{\frac{\pi}{4}{\rm sinc}^2(\delta)}\rightarrow1$ as $N\rightarrow\infty$, and $\hat{\mathbf{w}}_j^{RVQ}$ is the ZF precoding vector using the RVQ codebook for channel quantization.

Now, using a similar trick in \cite{1715541}, denote the quantization error $\alpha=\sin^2(\tilde{\mathbf{g}}_k,\hat{\mathbf{g}}_k^{RVQ})$ where $\hat{\mathbf{g}}_k^{RVQ}=\mathop{\argmax}_{\mathbf{d}_i\in\mathcal{G}^{RVQ}}|\tilde{\mathbf{g}}_k^H\mathbf{d}_i|$ in which $\mathcal{G}^{RVQ}=\{\mathbf{d}_{i}\}_{i=1}^{2^{B_2}}$ denotes the RVQ codebook. Mathmatically, we have $\tilde{\mathbf{g}}_k=\sqrt{1-\alpha}\hat{\mathbf{g}}_k^{RVQ}+\sqrt{\alpha} \mathbf{z}$ where $\mathbf{z}$ is a unit vector distributed in the null space of $\hat{\mathbf{g}}_k^{RVQ}$. Owing to the orthogonality between $\hat{\mathbf{g}}_k^{RVQ}$ and $\hat{\mathbf{w}}_j^{RVQ}$, we obtain
\begin{align}
|\tilde{\mathbf{g}}_k^H\hat{\mathbf{w}}_j^{RVQ}|^2=\alpha|\mathbf{z}^H\hat{\mathbf{w}}_j^{RVQ}|^2.\label{eq60}
\end{align}
Since $\mathbf{z}$ and $\hat{\mathbf{w}}_j^{RVQ}$ are independent, and both of them are isotropically distributed in the $(K-1)$-dimensional null-space of $\hat{\mathbf{g}}_k^{RVQ}$, we get \cite{7160780}
\begin{align}
\mathbb{E}\left[|\mathbf{z}^H\hat{\mathbf{w}}_j^{RVQ}|^2\right]=\frac{1}{K-1}.\label{eq61}
\end{align}

Then regarding $\alpha$, since $\hat{\mathbf{g}}_k^{RVQ}=\mathop{\argmax}_{\mathbf{d}_i\in\mathcal{G}^{RVQ}}|\tilde{\mathbf{g}}_k^H\mathbf{d}_i|$, we have $\mathrm{Pr}\{|\tilde{\mathbf{g}}_k^H\hat{\mathbf{g}}_k^{RVQ}|^2<y\}=\mathrm{Pr}\{|\tilde{\mathbf{g}}_k^H\mathbf{d}_{1}|^2<y,|\tilde{\mathbf{g}}_k^H\mathbf{d}_{2}|^2<y,...,|\tilde{\mathbf{g}}_k^H\mathbf{d}_{2^{B_2}}|^2<y\}\overset{(a)}{=}\left\{\mathrm{Pr}\{|\tilde{\mathbf{g}}_k^H\mathbf{d}_{i}|^2<y\}\right\}^{2^{B_2}}$ where (a) is true due to the independence among $\{|\tilde{\mathbf{g}}_k^H\mathbf{d}_{i}|^2\}_{i=1}^{2^{B_2}}$ \cite{7160780}. Defining $Y=1-\alpha=1-\sin^2(\tilde{\mathbf{g}}_k,\hat{\mathbf{g}}_k^{RVQ})=\cos^2(\tilde{\mathbf{g}}_k,\hat{\mathbf{g}}_k^{RVQ})=|\tilde{\mathbf{g}}_k^H\hat{\mathbf{g}}_k^{RVQ}|^2$, it yields
\begin{align}
\mathrm{Pr}\{Y<y\}=\left\{\mathrm{Pr}\{|\tilde{\mathbf{g}}_k^H\mathbf{d}_{i}|^2<y\}\right\}^{2^{B_2}}.\label{16}
\end{align}
Knowing that $g_{k,k}\overset{a.s.}{\rightarrow}{\rm sinc}\left(\frac{\pi}{2^{B_1}}\right)\frac{\sqrt{\pi}}{2}$ and $g_{k,j}\overset{a.s.}{\rightarrow}0$ from \eqref{eq51} and \eqref{eq52}
, we rewrite $|\tilde{\mathbf{g}}_k^H\mathbf{d}_{i}|^2$ as
\begin{align}
|\tilde{\mathbf{g}}_k^H\mathbf{d}_{i}|^2=\frac{|\mathbf{g}_k^H\mathbf{d}_{i}|^2}{\|\mathbf{g}_k\|^2\|\mathbf{d}_{i}\|^2}
=\left|\sum_{j=1}^Kg_{k,j}d_{i,j}\right|^2\frac{1}{(|\sum_{j=1}^Kg_{k,j}|^2)\|\mathbf{d}_{i}\|^2}
\overset{\overset{(a)}{a.s.}}{\rightarrow}|g_{k,k}d_{i,k}|^2\frac{1}{|g_{k,k}|^2\|\mathbf{d}_{i}\|^2}
=\frac{|d_{i,k}|^2}{\|\mathbf{d}_{i}\|^2}\label{17}
\end{align}
where $g_{k,j}$ is the $j$-th element of $\mathbf{g}_k$ and (a) utilizes the Continuous Mapping Theorem with continuous $\frac{1}{(|\sum_{j=1}^Kg_{k,j}|^2)\|\mathbf{d}_{i}\|^2}$ since $\|\mathbf{g}_k\|^2>0$ \cite{8287933} \cite{7953567}. 
From \cite{rodriguez2002quantum}, $\frac{d_{i,j}}{\|\mathbf{d}_{i}\|}$ could be expressed as
\begin{align}
\frac{d_{i,j}}{\|\mathbf{d}_{i}\|}=
\left\{\begin{array}{ll}
\cos\phi_1, &j=k\\
\sin\phi_1...\sin\phi_{K-2}\sin\phi_{K-1}, &j=K\\
\sin\phi_1...\sin\phi_{j}\cos\phi_{j+1}, &j<k\\
\sin\phi_1...\sin\phi_{j-1}\cos\phi_{j}, &{\rm otherwise}
\end{array}\right.\label{18}
\end{align}
where the angles $\phi_{1},\phi_{2},...,\phi_{K-2}$ range over $[0,\pi]$ and $\phi_{K-1}$ ranges over $[0,2\pi]$. Combining \eqref{17} and \eqref{18}, it yields, for $0<y<1$,
\begin{align}
\mathrm{\mathrm{Pr}}\{|\tilde{\mathbf{g}}_k^H\mathbf{d}_{i}|^2<y\}=&\mathrm{\mathrm{Pr}}\{\cos^2\phi_1<y\}
=\int_{\arccos\sqrt{y}}^{\pi-\arccos\sqrt{y}}\frac{1}{\pi}dy
=1-\frac{2}{\pi}\arccos\sqrt{y}
\overset{(a)}{\leq}1-(1-y)^{K-1}\label{19}
\end{align}
where inequality (a) can be acquired by analyzing the property of the function $h(y)\overset{\triangle}{=}[1-\frac{2}{\pi}\arccos\sqrt{y}]-[1-(1-y)^{K-1}]$. By checking the second-order derivative of $h(y)$, we know that it is a convex function for $y\in(0,1)$. Moreover, it is clear that $h(0)=h(1)=0$, we easily get that $h(y)\leq0$ holds for $y\in(0,1)$. Utilizing the fact that $\mathbb{E}[Y]=\int_0^1\mathrm{\mathrm{Pr}}\{Y\geq y\}dy=1-\int_0^1\mathrm{\mathrm{Pr}}\{Y<y\}dy$ for $0<y<1$ and $Y=1-\alpha$, we have $\mathbb{E}[\alpha]=1-\mathbb{E}[Y]=\int_0^1\mathrm{\mathrm{Pr}}\{Y<y\}dy$. Recalling $\mathrm{Pr}\{Y<y\}=\left\{\mathrm{Pr}\{|\tilde{\mathbf{g}}_k^H\mathbf{d}_{i}|^2<y\}\right\}^{2^{B_2}}$ in \eqref{16}, we further write
\begin{align}
\mathbb{E}[\alpha]=&\int_0^1\left\{\mathrm{Pr}\{|\tilde{\mathbf{g}}_k^H\mathbf{d}_{i}|^2<y\}\right\}^{2^{B_2}}dy\nonumber\\
\overset{(a)}{\leq}&\int_0^1(1-(1-y)^{K-1})^{2^{B_2}}dy\nonumber\\
\overset{(b)}{=}&\int_0^1(1-s^{K-1})^{2^{B_2}}ds\nonumber\\
\overset{(c)}{\leq}&2^{-\frac{B_2}{K-1}}\label{eq62}
\end{align}
where (a) uses \eqref{19}, (b) is obtained by setting $s=1-y$, and (c) is achieved based on \cite[Appendices I and II]{1715541}. From \eqref{eq60}, \eqref{eq61} and \eqref{eq62}, we have
\begin{align}
\mathbb{E}\left[|\tilde{\mathbf{g}}_k^H\hat{\mathbf{w}}_j^{RVQ}|^2\right]\leq\frac{1}{K-1}2^{-\frac{B_2}{K-1}}.\label{eq64}
\end{align}
Finally, substituting \eqref{eq64} into \eqref{eq63}, we achieve the first inequality in \eqref{eq31}.

\end{appendices}

\bibliographystyle{ieeetr}
\bibliography{myreference}

\end{document}